\documentclass[conference,letterpaper,nofonttune]{IEEEtran}  
\IEEEoverridecommandlockouts

\usepackage{amsmath,amssymb,amsfonts}
\usepackage[ruled,vlined]{algorithm2e}
\usepackage{graphicx}
\usepackage{textcomp}
\usepackage{xcolor}
\usepackage{soul}
\usepackage{balance}
\usepackage{subcaption}
\usepackage{glossaries}
\newacronym{3gpp}{3GPP}{3rd Generation Partnership Project}
\newacronym{4g}{4G}{4th generation mobile network}
\newacronym{5g}{5G}{5th generation mobile network}
\newacronym{6g}{6G}{6th generation mobile network}
\newacronym{nextg}{NextG}{Next Generation}
\newacronym{5gc}{5GC}{5G Core}
\newacronym{adc}{ADC}{Analog to Digital Converter}
\newacronym{aerpaw}{AERPAW}{Aerial Experimentation and Research Platform for Advanced Wireless}
\newacronym{ai}{AI}{Artificial Intelligence}
\newacronym{aimd}{AIMD}{Additive Increase Multiplicative Decrease}
\newacronym{am}{AM}{Acknowledged Mode}
\newacronym{amc}{AMC}{Adaptive Modulation and Coding}
\newacronym{amf}{AMF}{Access and Mobility Management Function}
\newacronym{aops}{AOPS}{Adaptive Order Prediction Scheduling}
\newacronym{api}{API}{Application Programming Interface}
\newacronym{apn}{APN}{Access Point Name}
\newacronym{aqm}{AQM}{Active Queue Management}
\newacronym{ausf}{AUSF}{Authentication Server Function}
\newacronym{avc}{AVC}{Advanced Video Coding}
\newacronym{awgn}{AGWN}{Additive White Gaussian Noise}
\newacronym{balia}{BALIA}{Balanced Link Adaptation Algorithm}
\newacronym{bbu}{BBU}{Base Band Unit}
\newacronym{bdp}{BDP}{Bandwidth-Delay Product}
\newacronym{ber}{BER}{Bit Error Rate}
\newacronym{bf}{BF}{Beamforming}
\newacronym{bler}{BLER}{Block Error Rate}
\newacronym{brr}{BRR}{Bayesian Ridge Regressor}
\newacronym{bsr}{BSR}{Buffer Status Report}
\newacronym{bs}{BS}{Base Station}
\newacronym{bpsk}{BPSK}{Binary Phase-shift keying}
\newacronym{bss}{BSS}{Business Support System}
\newacronym{ca}{CA}{Carrier Aggregation}
\newacronym{caas}{CaaS}{Connectivity-as-a-Service}
\newacronym{cb}{CB}{Code Block}
\newacronym{cc}{CC}{Congestion Control}
\newacronym{ccid}{CCID}{Congestion Control ID}
\newacronym{cco}{CC}{Carrier Component}
\newacronym{cd}{CD}{Continuous Delivery}
\newacronym{cdd}{CDD}{Cyclic Delay Diversity}
\newacronym{cdf}{CDF}{Cumulative Distribution Function}
\newacronym{cdma}{CDMA}{Code-Division Multiple Access}
\newacronym{cdn}{CDN}{Content Distribution Network}
\newacronym{ci}{CI}{Continuous Integration}
\newacronym{cicd}{CI/CD}{Continuous Integration/Continuous Delivery}
\newacronym{cir}{CIR}{Channel Impulse Response}
\newacronym{cn}{CN}{Core Network}
\newacronym{codel}{CoDel}{Controlled Delay Management}
\newacronym{comac}{COMAC}{Converged Multi-Access and Core}
\newacronym{cord}{CORD}{Central Office Re-architected as a Datacenter}
\newacronym{cornet}{CORNET}{COgnitive Radio NETwork}
\newacronym{cosmos}{COSMOS}{Cloud Enhanced Open Software Defined Mobile Wireless Testbed for City-Scale Deployment}
\newacronym{cots}{COTS}{Commercial Off-the-Shelf}
\newacronym{cp}{CP}{Control Plane}
\newacronym{cpu}{CPU}{Central Processing Unit}
\newacronym{cqi}{CQI}{Channel Quality Information}
\newacronym{cr}{CR}{Cognitive Radio}
\newacronym{cran}{CRAN}{Cloud \gls{ran}}
\newacronym{crs}{CRS}{Cell Reference Signal}
\newacronym{csi}{CSI}{Channel State Information}
\newacronym{csirs}{CSI-RS}{Channel State Information - Reference Signal}
\newacronym{cu}{CU}{Central Unit}
\newacronym{d2tcp}{D$^2$TCP}{Deadline-aware Data center TCP}
\newacronym{d3}{D$^3$}{Deadline-Driven Delivery}
\newacronym{dac}{DAC}{Digital to Analog Converter}
\newacronym{dag}{DAG}{Directed Acyclic Graph}
\newacronym{darpa}{DARPA}{Defense Advanced Research Projects Agency}
\newacronym{das}{DAS}{Distributed Antenna System}
\newacronym{dash}{DASH}{Dynamic Adaptive Streaming over HTTP}
\newacronym{dc}{DC}{Dual Connectivity}
\newacronym{dccp}{DCCP}{Datagram Congestion Control Protocol}
\newacronym{dce}{DCE}{Direct Code Execution}
\newacronym{dci}{DCI}{Downlink Control Information}
\newacronym{dcl}{DCL}{Dear Colleague Letter}
\newacronym{dctcp}{DCTCP}{Data Center TCP}
\newacronym{devops}{DevOps}{Development and Operations}
\newacronym{dl}{DL}{Downlink}
\newacronym{dmr}{DMR}{Deadline Miss Ratio}
\newacronym{dmrs}{DMRS}{DeModulation Reference Signal}
\newacronym{drlcc}{DRL-CC}{Deep Reinforcement Learning Congestion Control}
\newacronym{drs}{DRS}{Discovery Reference Signal}
\newacronym{dtn}{DTN}{Digital Twin Network}
\newacronym{dtmn}{DTMN}{Digital Twins for Mobile Networks}
\newacronym{dtwn}{DTWN}{Digital Twin Wireless Network}
\newacronym{du}{DU}{Distributed Unit}
\newacronym{e2e}{E2E}{end-to-end}
\newacronym{ecaas}{ECaaS}{Edge-Cloud-as-a-Service}
\newacronym{ecn}{ECN}{Explicit Congestion Notification}
\newacronym{edf}{EDF}{Earliest Deadline First}
\newacronym{em}{EM}{Electro-Magnetic}
\newacronym{embb}{eMBB}{Enhanced Mobile Broadband}
\newacronym{empower}{EMPOWER}{EMpowering transatlantic PlatfOrms for advanced WirEless Research}
\newacronym{enb}{eNB}{evolved Node Base}
\newacronym{endc}{EN-DC}{E-UTRAN-\gls{nr} \gls{dc}}
\newacronym{epc}{EPC}{Evolved Packet Core}
\newacronym{eps}{EPS}{Evolved Packet System}
\newacronym{es}{ES}{Edge Server}
\newacronym{etsi}{ETSI}{European Telecommunications Standards Institute}
\newacronym[firstplural=Estimated Times of Arrival (ETAs)]{eta}{ETA}{Estimated Time of Arrival}
\newacronym{eutran}{E-UTRAN}{Evolved Universal Terrestrial Access Network}
\newacronym{faas}{FaaS}{Function-as-a-Service}
\newacronym{fapi}{FAPI}{Functional Application Platform Interface}
\newacronym{fcc}{FCC}{Federal Communications Commission}
\newacronym{fdd}{FDD}{Frequency Division Duplexing}
\newacronym{fdm}{FDM}{Frequency Division Multiplexing}
\newacronym{fdma}{FDMA}{Frequency Division Multiple Access}
\newacronym{fed4fire}{FED4FIRE+}{Federation 4 Future Internet Research and Experimentation Plus}
\newacronym{fir}{FIR}{Finite Impulse Response}
\newacronym{fl}{FL}{Federated Learning}
\newacronym{fpga}{FPGA}{Field Programmable Gate Array}
\newacronym{fr2}{FR2}{Frequency Range 2}
\newacronym{fs}{FS}{Fast Switching}
\newacronym{fscc}{FSCC}{Flow Sharing Congestion Control}
\newacronym{ftp}{FTP}{File Transfer Protocol}
\newacronym{fw}{FW}{Flow Window}
\newacronym{ga128}{Ga}{Golay Sequence type A}
\newacronym{ge}{GE}{Gaussian Elimination}
\newacronym{glfsr}{GLFSR}{Galois Linear Feedback Shift Register}
\newacronym{gnb}{gNB}{Next Generation Node Base}
\newacronym{gold}{Gold}{Gold}
\newacronym{gop}{GOP}{Group of Pictures}
\newacronym{gpr}{GPR}{Gaussian Process Regressor}
\newacronym{gpu}{GPU}{Graphics Processing Unit}
\newacronym{gtp}{GTP}{GPRS Tunneling Protocol}
\newacronym{gtpc}{GTP-C}{GPRS Tunnelling Protocol Control Plane}
\newacronym{gtpu}{GTP-U}{GPRS Tunnelling Protocol User Plane}
\newacronym{gtpv2c}{GTPv2-C}{\gls{gtp} v2 - Control}
\newacronym{gw}{GW}{Gateway}
\newacronym{harq}{HARQ}{Hybrid Automatic Repeat reQuest}
\newacronym{hetnet}{HetNet}{Heterogeneous Network}
\newacronym{hh}{HH}{Hard Handover}
\newacronym{hol}{HOL}{Head-of-Line}
\newacronym{hqf}{HQF}{Highest-quality-first}
\newacronym{hss}{HSS}{Home Subscription Server}
\newacronym{http}{HTTP}{HyperText Transfer Protocol}
\newacronym{ia}{IA}{Initial Access}
\newacronym{iab}{IAB}{Integrated Access and Backhaul}
\newacronym{ic}{IC}{Incident Command}
\newacronym{ietf}{IETF}{Internet Engineering Task Force}
\newacronym{ifw}{IFW}{Interference Free Window}
\newacronym{imsi}{IMSI}{International Mobile Subscriber Identity}
\newacronym{imt}{IMT}{International Mobile Telecommunication}
\newacronym{iot}{IoT}{Internet of Things}
\newacronym{ip}{IP}{Internet Protocol}
\newacronym{iq}{IQ}{In-phase and Quadrature}
\newacronym{isi}{ISI}{Inter-Symbol Interference}
\newacronym{itu}{ITU}{International Telecommunication Union}
\newacronym{kpi}{KPI}{Key Performance Indicator}
\newacronym{kvm}{KVM}{Kernel-based Virtual Machine}
\newacronym{lfsr}{LFSR}{Linear Feedback Shift Register}
\newacronym{los}{LOS}{Line-of-Sight}
\newacronym{ls}{LS}{Loosely Synchronised}
\newacronym{lsm}{LSM}{Link-to-System Mapping}
\newacronym{lstm}{LSTM}{Long Short Term Memory}
\newacronym{lte}{LTE}{Long Term Evolution}
\newacronym{lxc}{LXC}{Linux Container}
\newacronym{m2m}{M2M}{Machine to Machine}
\newacronym{mac}{MAC}{Medium Access Control}
\newacronym{mai}{MAI}{Multiple Access Interference}
\newacronym{manet}{MANET}{Mobile Ad Hoc Network}
\newacronym{mano}{MANO}{Management and Orchestration}
\newacronym{mc}{MC}{Multi-Connectivity}
\newacronym{mcc}{MCC}{Mobile Cloud Computing}
\newacronym{mchem}{MCHEM}{Massive Channel Emulator}
\newacronym{mcs}{MCS}{Modulation and Coding Scheme}
\newacronym{mec}{MEC}{Multi-access Edge Computing}
\newacronym{mec2}{MEC}{Mobile Edge Cloud}
\newacronym{mec3}{MEC}{Mobile Edge Computing}
\newacronym{mfc}{MFC}{Mobile Fog Computing}
\newacronym{mi}{MI}{Mutual Information}
\newacronym{mib}{MIB}{Master Information Block}
\newacronym{miesm}{MIESM}{Mutual Information Based Effective SINR}
\newacronym{mimo}{MIMO}{Multiple Input, Multiple Output}
\newacronym{mgen}{MGEN}{Multi-Generator}
\newacronym{ml}{ML}{Machine Learning}
\newacronym{mlr}{MLR}{Maximum-local-rate}
\newacronym[plural=\gls{mme}s,firstplural=Mobility Management Entities (MMEs)]{mme}{MME}{Mobility Management Entity}
\newacronym{mmtc}{mMTC}{Massive Machine-Type Communications}
\newacronym{mmwave}{mmWave}{millimeter wave}
\newacronym{mpdccp}{MP-DCCP}{Multipath Datagram Congestion Control Protocol}
\newacronym{mptcp}{MPTCP}{Multipath TCP}
\newacronym{mr}{MR}{Maximum Rate}
\newacronym{mrdc}{MR-DC}{Multi \gls{rat} \gls{dc}}
\newacronym{mse}{MSE}{Mean Square Error}
\newacronym{mss}{MSS}{Maximum Segment Size}
\newacronym{mt}{MT}{Mobile Termination}
\newacronym{mtd}{MTD}{Machine-Type Device}
\newacronym{mtu}{MTU}{Maximum Transmission Unit}
\newacronym{mumimo}{MU-MIMO}{Multi-user \gls{mimo}}
\newacronym{mvno}{MVNO}{Mobile Virtual Network Operator}
\newacronym{nalu}{NALU}{Network Abstraction Layer Unit}
\newacronym{nas}{NAS}{Network Attached Storage}
\newacronym{nbiot}{NB-IoT}{Narrow Band IoT}
\newacronym{nfv}{NFV}{Network Function Virtualization}
\newacronym{nfvi}{NFVI}{Network Function Virtualization Infrastructure}
\newacronym{nic}{NIC}{Network Interface Card}
\newacronym{nlos}{NLOS}{Non-Line-of-Sight}
\newacronym{now}{NOW}{Non Overlapping Window}
\newacronym{nrdz}{NRDZ}{National Radio Dynamic Zone}
\newacronym{nsf}{NSF}{National Science Foundation}
\newacronym{nsm}{NSM}{Network Service Mesh}
\newacronym{nr}{NR}{New Radio}
\newacronym{nrf}{NRF}{Network Repository Function}
\newacronym{nsa}{NSA}{Non Stand Alone}
\newacronym{nse}{NSE}{Network Slicing Engine}
\newacronym{nssf}{NSSF}{Network Slice Selection Function}
\newacronym{ntp}{NTP}{Network Time Protocol}
\newacronym{o2i}{O2I}{Outdoor to Indoor}
\newacronym{oai}{OAI}{OpenAirInterface}
\newacronym{oaicn}{OAI-CN}{\gls{oai} \acrlong{cn}}
\newacronym{oairan}{OAI-RAN}{\acrlong{oai} \acrlong{ran}}
\newacronym{oam}{OAM}{Operations, Administration and Maintenance}
\newacronym[plural=\gls{obu}s,firstplural=Onboard Units (OBUs)]{obu}{OBU}{Onboard Unit}
\newacronym{ofdm}{OFDM}{Orthogonal Frequency Division Multiplexing}
\newacronym{olia}{OLIA}{Opportunistic Linked Increase Algorithm}
\newacronym{omec}{OMEC}{Open Mobile Evolved Core}
\newacronym{onap}{ONAP}{Open Network Automation Platform}
\newacronym{onf}{ONF}{Open Networking Foundation}
\newacronym{onos}{ONOS}{Open Networking Operating System}
\newacronym{oom}{OOM}{\gls{onap} Operations Manager}
\newacronym{opnfv}{OPNFV}{Open Platform for \gls{nfv}}
\newacronym{orbit}{ORBIT}{Open-Access Research Testbed for Next-Generation Wireless Networks}
\newacronym{os}{OS}{Operating System}
\newacronym{osm}{OSM}{Open Street Map}
\newacronym{oss}{OSS}{Operations Support System}
\newacronym{pa}{PA}{Position-aware}
\newacronym{pase}{PASE}{Prioritization, Arbitration, and Self-adjusting Endpoints}
\newacronym{pawr}{PAWR}{Platforms for Advanced Wireless Research}
\newacronym{pbch}{PBCH}{Physical Broadcast Channel}
\newacronym{pcef}{PCEF}{Policy and Charging Enforcement Function}
\newacronym{pcfich}{PCFICH}{Physical Control Format Indicator Channel}
\newacronym{pcrf}{PCRF}{Policy and Charging Rules Function}
\newacronym{pdcch}{PDCCH}{Physical Downlink Control Channel}
\newacronym{pdcp}{PDCP}{Packet Data Convergence Protocol}
\newacronym{pdsch}{PDSCH}{Physical Downlink Shared Channel}
\newacronym{pdu}{PDU}{Packet Data Unit}
\newacronym{pdp}{PDP}{Power Delay Profile}
\newacronym{pf}{PF}{Proportional Fair}
\newacronym{pgw}{PGW}{Packet Gateway}
\newacronym{phich}{PHICH}{Physical Hybrid ARQ Indicator Channel}
\newacronym{phy}{PHY}{Physical}
\newacronym{pl}{PL}{Path Loss}
\newacronym{pmch}{PMCH}{Physical Multicast Channel}
\newacronym{pmi}{PMI}{Precoding Matrix Indicators}
\newacronym{powder}{POWDER}{Platform for Open Wireless Data-driven Experimental Research}
\newacronym{ppo}{PPO}{Proximal Policy Optimization}
\newacronym{ppp}{PPP}{Poisson Point Process}
\newacronym{prach}{PRACH}{Physical Random Access Channel}
\newacronym{prb}{PRB}{Physical Resource Block}
\newacronym{psnr}{PSNR}{Peak Signal to Noise Ratio}
\newacronym{pss}{PSS}{Primary Synchronization Signal}
\newacronym{pucch}{PUCCH}{Physical Uplink Control Channel}
\newacronym{pusch}{PUSCH}{Physical Uplink Shared Channel}
\newacronym{qam}{QAM}{Quadrature Amplitude Modulation}
\newacronym{qci}{QCI}{\gls{qos} Class Identifier}
\newacronym{qoe}{QoE}{Quality of Experience}
\newacronym{qos}{QoS}{Quality of Service}
\newacronym{qtgui}{QT-GUI}{QT Graphical User Interface}
\newacronym{quic}{QUIC}{Quick UDP Internet Connections}
\newacronym{rach}{RACH}{Random Access Channel}
\newacronym{ran}{RAN}{Radio Access Network}
\newacronym[firstplural=Radio Access Technologies (RATs)]{rat}{RAT}{Radio Access Technology}
\newacronym{rcn}{RCN}{Research Coordination Network}
\newacronym{rec}{REC}{Radio Edge Cloud}
\newacronym{red}{RED}{Random Early Detection}
\newacronym{renew}{RENEW}{Reconfigurable Eco-system for Next-generation End-to-end Wireless}
\newacronym{rf}{RF}{Radio Frequency}
\newacronym{rfc}{RFC}{Request for Comments}
\newacronym{rfr}{RFR}{Random Forest Regressor}
\newacronym{ric}{RIC}{\gls{ran} Intelligent Controller}
\newacronym{rlc}{RLC}{Radio Link Control}
\newacronym{rlf}{RLF}{Radio Link Failure}
\newacronym{rlnc}{RLNC}{Random Linear Network Coding}
\newacronym{rmse}{RMSE}{Root Mean Squared Error}
\newacronym{rnis}{RNIS}{Radio Network Information Service}
\newacronym{rr}{RR}{Round Robin}
\newacronym{rrc}{RRC}{Radio Resource Control}
\newacronym{rrm}{RRM}{Radio Resource Management}
\newacronym{rru}{RRU}{Remote Radio Unit}
\newacronym{rs}{RS}{Remote Server}
\newacronym{rsrp}{RSRP}{Reference Signal Received Power}
\newacronym{rsrq}{RSRQ}{Reference Signal Received Quality}
\newacronym{rss}{RSS}{Received Signal Strength}
\newacronym{rssi}{RSSI}{Received Signal Strength Indicator}
\newacronym{rsu}{RSU}{Road-Side Unit}
\newacronym{rtt}{RTT}{Round Trip Time}
\newacronym{ru}{RU}{Radio Unit}
\newacronym{rw}{RW}{Receive Window}
\newacronym{rx}{RX}{Receiver}
\newacronym{s1ap}{S1AP}{S1 Application Protocol}
\newacronym{sa}{SA}{standalone}
\newacronym{sack}{SACK}{Selective Acknowledgment}
\newacronym{sap}{SAP}{Service Access Point}
\newacronym{sc2}{SC2}{Spectrum Collaboration Challenge}
\newacronym{scef}{SCEF}{Service Capability Exposure Function}
\newacronym{sch}{SCH}{Secondary Cell Handover}
\newacronym{scoot}{SCOOT}{Split Cycle Offset Optimization Technique}
\newacronym{sctp}{SCTP}{Stream Control Transmission Protocol}
\newacronym{sdap}{SDAP}{Service Data Adaptation Protocol}
\newacronym{sd}{SD}{Standard Deviation}
\newacronym{sdk}{SDK}{Software Development Kit}
\newacronym{sdm}{SDM}{Space Division Multiplexing}
\newacronym{sdma}{SDMA}{Spatial Division Multiple Access}
\newacronym{sdn}{SDN}{Software-defined Networking}
\newacronym{sdr}{SDR}{Software-defined Radio}
\newacronym{seba}{SEBA}{SDN-Enabled Broadband Access}
\newacronym{sgsn}{SGSN}{Serving GPRS Support Node}
\newacronym{sgw}{SGW}{Service Gateway}
\newacronym{si}{SI}{Study Item}
\newacronym{sib}{SIB}{Secondary Information Block}
\newacronym{sinr}{SINR}{Signal to Interference plus Noise Ratio}
\newacronym{sip}{SIP}{Session Initiation Protocol}
\newacronym{siso}{SISO}{Single Input, Single Output}
\newacronym{sla}{SLA}{Service Level Agreement}
\newacronym{sm}{SM}{Saturation Mode}
\newacronym{smf}{SMF}{Session Management Function}
\newacronym{smo}{SMO}{Service Management and Orchestration}
\newacronym{sms}{SMS}{Short Message Service}
\newacronym{smsgmsc}{SMS-GMSC}{\gls{sms}-Gateway}
\newacronym{snr}{SNR}{Signal-to-Noise-Ratio}
\newacronym{son}{SON}{Self-Organizing Network}
\newacronym{sptcp}{SPTCP}{Single Path TCP}
\newacronym{srb}{SRB}{Service Radio Bearer}
\newacronym{srn}{SRN}{Standard Radio Node}
\newacronym{srs}{SRS}{Sounding Reference Signal}
\newacronym{ss}{SS}{Synchronization Signal}
\newacronym{sss}{SSS}{Secondary Synchronization Signal}
\newacronym{st}{ST}{Spanning Tree}
\newacronym{svc}{SVC}{Scalable Video Coding}
\newacronym{tb}{TB}{Transport Block}
\newacronym{tcp}{TCP}{Transmission Control Protocol}
\newacronym{tdd}{TDD}{Time Division Duplexing}
\newacronym{tdm}{TDM}{Time Division Multiplexing}
\newacronym{tdma}{TDMA}{Time Division Multiple Access}
\newacronym{tfl}{TfL}{Transport for London}
\newacronym{tfrc}{TFRC}{TCP-Friendly Rate Control}
\newacronym{tft}{TFT}{Traffic Flow Template}
\newacronym{tgen}{TGEN}{Traffic Generator}
\newacronym{tip}{TIP}{Telecom Infra Project}
\newacronym{tm}{TM}{Transparent Mode}
\newacronym{to}{TO}{Telco Operator}
\newacronym{toa}{ToA}{Time of Arrival}
\newacronym{tr}{TR}{Technical Report}
\newacronym{trp}{TRP}{Transmitter Receiver Pair}
\newacronym{ts}{TS}{Technical Specification}
\newacronym{tti}{TTI}{Transmission Time Interval}
\newacronym{ttt}{TTT}{Time-to-Trigger}
\newacronym{tx}{TX}{Transmitter}
\newacronym{uas}{UAS}{Unmanned Aerial System}
\newacronym{uav}{UAV}{Unmanned Aerial Vehicle}
\newacronym{udm}{UDM}{Unified Data Management}
\newacronym{udp}{UDP}{User Datagram Protocol}
\newacronym{udr}{UDR}{Unified Data Repository}
\newacronym{ue}{UE}{User Equipment}
\newacronym{uhd}{UHD}{\gls{usrp} Hardware Driver}
\newacronym{ul}{UL}{Uplink}
\newacronym{um}{UM}{Unacknowledged Mode}
\newacronym{uml}{UML}{Unified Modeling Language}
\newacronym{upa}{UPA}{Uniform Planar Array}
\newacronym{upf}{UPF}{User Plane Function}
\newacronym{urllc}{URLLC}{Ultra Reliable and Low Latency Communication}
\newacronym{usa}{U.S.}{United States}
\newacronym{usim}{USIM}{Universal Subscriber Identity Module}
\newacronym{usrp}{USRP}{Universal Software Radio Peripheral}
\newacronym{utc}{UTC}{Urban Traffic Control}
\newacronym{vim}{VIM}{Virtualization Infrastructure Manager}
\newacronym{vm}{VM}{Virtual Machine}
\newacronym{vnf}{VNF}{Virtual Network Function}
\newacronym{volte}{VoLTE}{Voice over \gls{lte}}
\newacronym{voltha}{VOLTHA}{Virtual OLT HArdware Abstraction}
\newacronym{vr}{VR}{Virtual Reality}
\newacronym{vran}{vRAN}{Virtualized \gls{ran}}
\newacronym{vss}{VSS}{Video Streaming Server}
\newacronym{wbf}{WBF}{Wired Bias Function}
\newacronym{wf}{WF}{Wired-first}
\newacronym{wi}{WI}{Wireless InSite}
\newacronym{wlan}{WLAN}{Wireless Local Area Network}
\newacronym{pnf}{PNF}{Physical Network Function}
\newacronym{drl}{DRL}{Deep Reinforcement Learning}
\newacronym{mtc}{MTC}{Machine-type Communications}
\newacronym{v2x}{V2X}{Vehicle-to-everything}
\newacronym{cast}{CaST}{Channel emulation scenario generator and Sounder Toolchain}
\newacronym{gui}{GUI}{Graphical User Interface}
\newacronym{ups}{UPS}{Uninterruptible Power Supply}
\newacronym{ota}{OTA}{Over-the-Air}
\newacronym{hitl}{HITL}{hardware-in-the-loop}
\newacronym{soc}{SoC}{System-on-Chip}

\newacronym{psr}{PSR}{Packet Success Rate}
\newacronym{sadr}{SADR}{Safe Adaptive Data Rate}
\newacronym{cnn}{CNN}{Convolutional Neural Network}
\newacronym{kpm}{KPM}{Key Performance Measurement}
\newacronym{mqtt}{MQTT}{Message Queuing Telemetry Transport}
\newacronym{coap}{CoAP}{Constrained Application Protocol}
\newacronym{amqp}{AMQP}{Advanced Message Queuing Protocol}
\newacronym{dt}{DT}{Digital Twin}
\newacronym{dlai}{DL}{Deep Learning}

\newacronym{pla}{PLA}{Physical Layer Authentication}
\newacronym{pls}{PLS}{Physical Layer Security}
\newacronym{gmm}{GMM}{Gaussian Mixture Model}

\newacronym{sit}{SIT}{Signature Injection Tool}
\newacronym{dpi}{DPI}{Deep Packet Inspector}
\newacronym{ks}{K-S}{Kolmogorov-Smirnov}

\newacronym{dnn}{DNN}{Deep Neural Network}
\newacronym{pdf}{PDF}{Power Spectral Density}

\def\BibTeX{{\rm B\kern-.05em{\sc i\kern-.025em b}\kern-.08em T\kern-.1667em\lower.7ex\hbox{E}\kern-.125emX}}

\usepackage[numbers,sort&compress]{natbib}
\usepackage{booktabs} 
\usepackage{tabularx} 

\usepackage{tikz}
\usepackage{tikzscale}

\newif\ifexttikz
\exttikzfalse
\ifexttikz
\else
\usepackage{tikzpagenodes,etoolbox}
\usetikzlibrary{calc}
\usepackage[contents={}]{background}
\AddEverypageHook{%
\ifnumequal{\thepage}{1}{%
    \tikz[remember picture,overlay]{%
        \node[draw,
        minimum width=1.03\textwidth,
        text width=1.02\textwidth,
        font=\footnotesize
        ]
        at ($(current page header area) - (0,5pt)$)
        {%
        This paper has been accepted for publication on IEEE Military Communications Conference (MILCOM 2025). This is the author's accepted version of the article. The final version published by IEEE is CP. Robinson, S. D'Oro, and T. Melodia, ``\textit{\veriphy:} Physical Layer Signal Authentication for Wireless Communication in 5G Environments,'' Proc. of  \textit{IEEE Military Communications Conference (MILCOM)}, Los Angeles, CA, USA, 2025.
        };
        \node[draw,
        minimum width=1.03\textwidth,
        text width=1.02\textwidth,
        font=\footnotesize
        ]
        at (current page footer area)
        {%
        ©2025 IEEE. Personal use of this material is permitted. Permission from IEEE must be obtained for all other uses, in any current or future media, including reprinting/republishing this material for advertising or promotional purposes, creating new collective works, for resale or redistribution to servers or lists, or reuse of any copyrighted component of this work in other works.
        };
    }%
}{}
}
\fi

\newcommand{\veriphy}{\textit{VeriPHY}\xspace}

\begin{document}

\title{\textit{VeriPHY}: Physical Layer Signal Authentication for Wireless Communication in 5G Environments}

\author{
\IEEEauthorblockN{Clifton Paul Robinson, Salvatore D'Oro, and Tommaso Melodia\vspace{-0.3cm}}
\IEEEauthorblockN{Institute for the Wireless Internet of Things, Northeastern University, Boston, MA, USA\\
Email: \{cl.robinson, s.doro, t.melodia\}@northeastern.edu
}
\thanks{This work was partially funded by the U.S. National Science Foundation under grants CNS-1925601 and CNS-2112471.}
}


\maketitle

\begin{abstract}
Physical layer authentication (PLA) uses inherent characteristics of the communication medium to provide secure and efficient authentication in wireless networks, bypassing the need for traditional cryptographic methods. With advancements in deep learning, PLA has become a widely adopted technique for its accuracy and reliability. In this paper, we introduce \textit{VeriPHY}, a novel deep learning-based PLA solution for 5G networks, which enables unique device identification by embedding signatures within wireless I/Q transmissions using steganography. \textit{VeriPHY} continuously generates pseudo-random signatures by sampling from Gaussian Mixture Models whose distribution is carefully varied to ensure signature uniqueness and stealthiness over time, and then embeds the newly generated signatures over I/Q samples transmitted by users to the 5G gNB. Utilizing deep neural networks, \textit{VeriPHY} identifies and authenticates users based on these embedded signatures. \textit{VeriPHY} achieves high precision, identifying unique signatures between 93\% and 100\% with low false positive rates and an inference time of 28~ms when signatures are updated every 20~ms. Additionally, we also demonstrate a stealth generation mode where signatures are generated in a way that makes them virtually indistinguishable from unaltered 5G signals while maintaining over 93\% detection accuracy.
\end{abstract}


\section{Introduction}
\label{sec:intro}


\gls{pls} is crucial in modern communication systems for securing information starting from the RF domain. Within \gls{pls}, \gls{pla} is an important component to verify the legitimacy of communication entities based on their unique physical characteristics~\cite{bai_physical_2020}. As wireless networks and the number of devices continue to grow, developing robust \gls{pla} mechanisms is increasingly vital to combat unauthorized access and maintain data integrity~\cite{pla_intro_1}. 
Various \gls{pla} solutions have been developed over the years~\cite{pla_survey}, including challenge-response protocols based on physical layer parameters, fingerprinting based on hardware imperfections, and RFID-based methods using unique tag responses
. These methods form the foundation for advanced \gls{pla} in modern communication systems, which ideally exhibit three key characteristics~\cite{pla_intro_1}: \textbf{covertness}, ensuring undetectability by unauthorized entities; \textbf{robustness}, maintaining functionality in adverse conditions; and \textbf{security} against breaches.

As \glspl{5g} expand to support a vast ecosystem of devices and applications, \gls{pla} becomes critical for ensuring secure, reliable communication~\cite{pla_survey}. By verifying device identity and integrity at the physical layer, \gls{pla} mitigates risks from unauthorized access, cyber threats, and misuse of limited radio resources, thereby safeguarding 5G infrastructure and users~\cite{pla_survey}.

Effective \gls{pla} designs must address challenges such as accurate device identification, spoofing resistance, and real-time processing. The rise of \gls{dlai}, particularly \glspl{cnn}, has advanced the field by enabling adaptive, robust, and efficient authentication~\cite{alhoraibi_physical_2023}. \glspl{cnn} enhance \gls{pla} by distinguishing legitimate from illegitimate signals, adapting to dynamic conditions, and handling complex data with high accuracy. For example, PAST-AI~\cite{oligeri_past-ai_2023} used \glspl{cnn} to authenticate satellite transducers with up to 100\% accuracy, while~\cite{fi14020061} demonstrated effective node authentication and spoofing detection in wireless sensor networks. These approaches highlight the potential of \gls{dlai}-based methods for practical, real-world \gls{pla} deployment. 

\begin{figure}[!t]
    \centering
    \includegraphics[width=.95\linewidth]{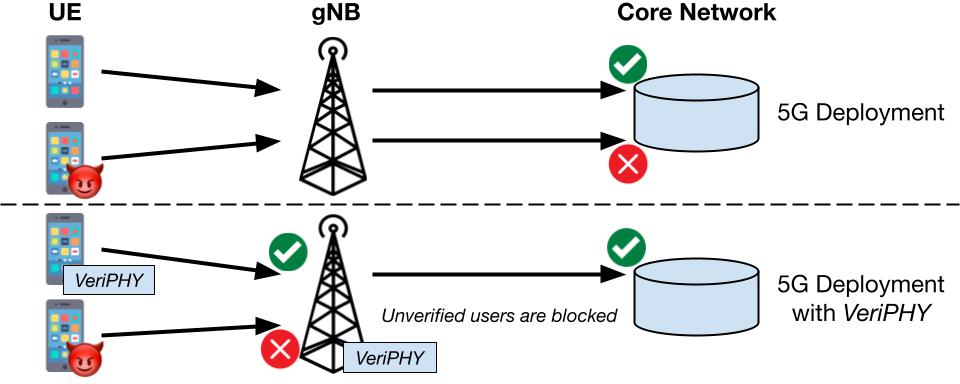}
    \caption{Comparison of authentication in traditional 5G networks (top) vs. \veriphy (bottom), which blocks unauthorized core access at the physical layer.} 
    \label{fig:veriphy_intro}
    \vspace{-0.3cm}
\end{figure}


To achieve covertness in \gls{pla} systems, steganography can be used as a way to hide authentication-related data and procedures behind overt data, i.e., data that any node can eavesdrop and decode, such as wireless signals~\cite{pla_survey}. This approach enhances security by embedding authentication messages within intelligible transmissions that only apparently do not carry any security-related data, thus thwarting interception or tampering~\cite{bonati_stealte_2021}.  
The use of wireless signatures~\cite{wang_physical-layer_2016, bai_physical_2020}, on the other hand, is another approach that relies on unique signal characteristics like variations in strength, timing, or frequency response to authenticate devices to create unique wireless fingerprints. Combining steganography with wireless signatures fortifies \gls{pla} systems with dual-layer security through covert data embedding and distinct signal attributes.

\begin{figure*}[!t]
    \vspace{0.1cm}
    \centering
    \includegraphics[width=.9\textwidth]{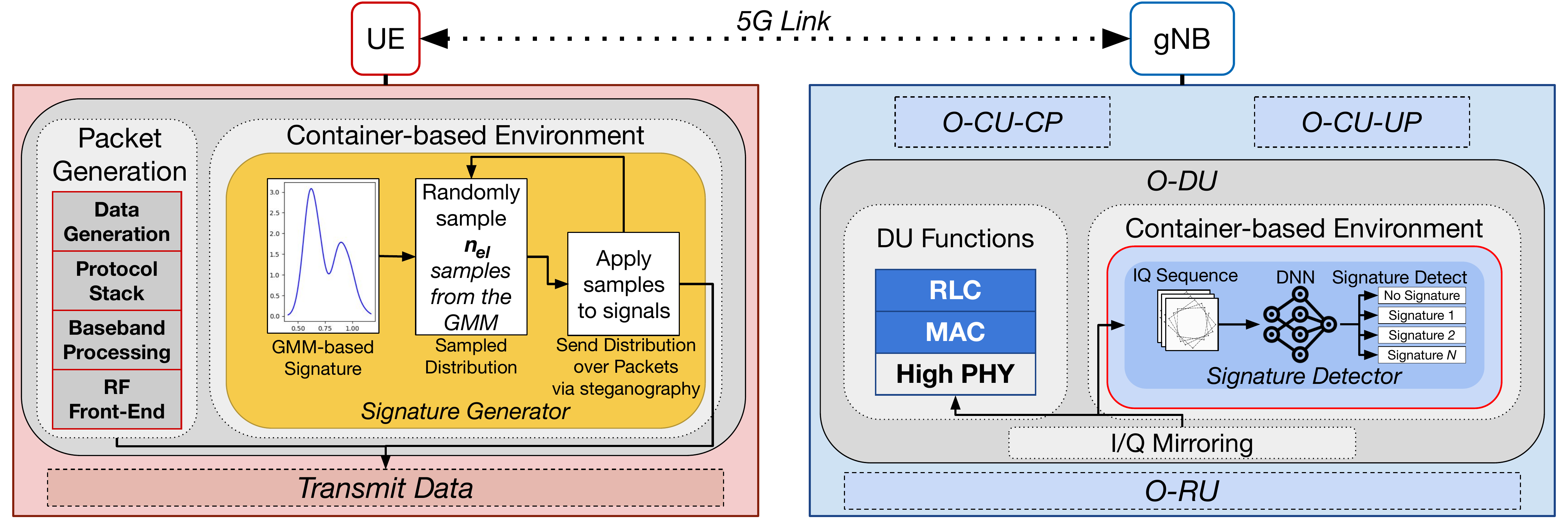}
    \caption{The high-level design \veriphy in both the \gls{ue} and gNB describing how the \gls{ue} generates the signatures to transmit with their packets and how the gNB uses I/Q mirroring to process and detect potential user signatures continuously throughout communication transmissions.}
    \label{fig:veriphy_main}
    \vspace{-0.3cm}
\end{figure*}

\glsreset{pla}
In this paper, we present \veriphy, a novel \gls{pla} solution for \gls{5g} that generates continuously pseudo-random and unique device signatures via steganography in wireless I/Q transmissions. \veriphy uses \gls{dlai}-based \glspl{cnn} to analyze I/Q samples in real time, allowing accurate physical layer authentication and blocking unauthorized users before their attachment requests reach the core network. As shown in Fig.~\ref{fig:veriphy_intro}, unlike standard 5G procedures that detect unauthorized users at the core, \veriphy intercepts them at the \gls{gnb}, enhancing early threat mitigation. Its embedded \gls{pls} mechanisms resist identity spoofing, preventing even advanced attackers (e.g., those cloning IMSIs) from replicating device signatures. This strengthens security by ensuring only legitimate devices access 5G resources.

\textbf{Novel contributions:} the important technical contributions of the paper can be summarized as follows:
\begin{enumerate}
    \item We introduce \veriphy, a novel \gls{dlai}-based 5G \gls{pla} solution that enables multi-UE identification by embedding wireless signatures into I/Qs using steganography;
    \item We utilize \glspl{gmm} to generate pseudo-random unique signatures for each device that cannot be replicated or forged. This randomization guarantees signature uniqueness and enhances the overall security of the \veriphy system;
    \item We demonstrate that our trained \glspl{dnn} can accurately detect unique signatures with 93\%-100\% accuracy and low false positives, achieving detection times as low as 6.5~ms when signatures are sent every 1~ms (below the 5G NR frame duration of 10~ms) and 25.5~ms when sent every 20~ms;
    \item We also introduce a \textbf{\textit{stealth mode}}, where \veriphy’s signatures are altered via a pre-processing function to be practically indistinguishable from standard 5G signals, yet remain detectable by \glspl{dnn} with over 93\% accuracy.
\end{enumerate}




\glsreset{pla}
\glsreset{gmm}
\section{Related Work}
\label{sec:relate}

\gls{pla} has emerged as a key technology for enhancing wireless security in \gls{5g} by leveraging physical layer attributes, addressing current limitations such as susceptibility to spoofing and unauthorized access, and integrating with existing infrastructure~\cite{wang_physical-layer_2016}.
%
%
%
\gls{pla} and security offer a promising approach to safeguard wireless communications by leveraging the randomness and stochasticity of channels, serving as an alternative to complex cryptographic techniques~\cite{wang_physical-layer_2016}. 
%


A key aspect of \gls{pla} is authenticating users and devices via unique physical-layer signatures. \cite{liu_authenticating_2010} proposes a method using helper nodes with channel-derived cryptographic signatures for authentication.
\textit{SteaLTE}~\cite{bonati_stealte_2021} applies wireless steganography to embed data covertly in cellular traffic without degrading performance.
In \cite{qiu_physical_2018}, probabilistic modeling of the channel and \glspl{gmm} are used to detect spoofing attacks, showing improved detection accuracy.


%
In \cite{liu_authenticating_2010}, the authors introduce an FCC-compliant authentication method for primary users that combines cryptographic and wireless link signatures, using a nearby helper node to authenticate signals without training. 
\cite{gulati_gmm_2013} presents a \gls{gmm}-based semi-supervised technique for channel-based authentication, achieving high detection with low false alarms and adapting to network changes without prior intruder data. 
\cite{scalingi2024det} proposes a real-time anomaly detection framework for 5G RRC-layer vulnerabilities, leveraging \gls{ai} to analyze PHY and cross-layer features. Validated in emulated and real environments, it achieved over 85\% detection accuracy with low latency, making it suitable for Open RAN deployment.


In the context of \gls{pla} and \gls{pls}, deep learning has become a popular method in classifying the wireless spectrum through \glspl{dnn}.
In \cite{bao_threat_2022}, the authors investigate attacks on \gls{cnn}-based device identification, proposing evaluation indicators to improve assessment. Higher perturbation levels and iteration steps degrade accuracy, providing insights for resilient \gls{dlai}-based \gls{iot} systems.
\cite{zhang_signal_2021} proposes a \gls{dnn} approach using \glspl{cnn} and \glspl{dnn} to classify multiple signals in shared-spectrum networks using I/Q samples, validated experimentally with USRP radios.
\cite{riyaz_deep_2020-1} proposes an intrusion detection system for wireless networks, employing feature selection algorithms with conditional random fields and linear correlation coefficients, integrated with \glspl{cnn} for classification, achieving a validated 99\% detection accuracy via tenfold cross-validation.


\veriphy differs from the literature above as it introduces a novel approach for device authentication via physical layer steganography, embedding ever changing signatures directly into transmitted packets. Unlike \cite{bonati_stealte_2021}, which uses steganography at the application layer, and \cite{liu_authenticating_2010}, which requires an additional node, \veriphy streamlines authentication between \glspl{ue} and \glspl{gnb} using unique user signatures at the physical layer, which can be hidden using a stealth mode activated by network operators. This advancement integrates \gls{pla} into \gls{5g} networks, highlighting the importance of high-accuracy signature authentication that can be concealed from eavesdroppers.

\section{\textit{VeriPHY} Signature Framework}
\label{sec:frame}

\veriphy's architecture is illustrated in Fig.~\ref{fig:veriphy_main}. Due to space limitations, in Fig.~\ref{fig:veriphy_main} and in the following we focus on the case where signatures are generated by \glspl{ue} and retrieved at the \gls{gnb}. However, we would like to mention that \veriphy is designed to support the execution of signature detection (left) and generation (right) modules at both 5G \glspl{ue} and \gls{gnb}, thus providing a framework for mutual authentication between both parties. 
In the above case, \veriphy uses two container-based environments: the Signature Generator in the \gls{ue} for generating user signatures, and the Signature Detector in the \gls{gnb}, which uses a trained \gls{dnn} to identify these signatures from I/Q samples. 
Thanks to the cloud-native design, the Signature Detector and Generator can be deployed at both the \gls{ue} and \gls{gnb} as software modules running as microservices and containers. This enhances flexibility and scalability, simplifying deployment, management, and integration in diverse environments.


The following subsections detail both components: Subsection~\ref{subsec:sig_gen} covers the Signature Generator, including how it creates and sends signatures, while Subsection~\ref{subsec:sig_detect} explains how the Signature Detector uses \gls{dlai} to detect and verify them.


\vspace{-0.1cm}
\glsreset{ue}
\subsection{Signature Generator}
\label{subsec:sig_gen}

The Signature Generator creates and stores unique user signatures essential for secure transmission and authentication. Embedded via steganography in I/Q data, these signatures form patterns detectable by allies but difficult for adversaries to replicate or detect. This enhances security across the \gls{5g} transceiver chain without disrupting normal operations, enabling fast detection and robust protection.

In a \veriphy-enabled \gls{5g} deployment, each \gls{ue} holds a unique \gls{gmm}-based signature distribution. A signature is generated by sampling $n_{el}$ samples from the \gls{ue}-specific \gls{gmm} distribution and transmitted every $t$~ms by embedding it on top of user data via wireless steganography as we will describe in detail in Section~\ref{sec:sig_gen}. Since signatures continuously change every few milliseconds (e.g., our prototype updates signatures every 20ms, i.e., 2 5G frames) by randomly sampling from the originating \gls{gmm}, it is hard for attackers to replicate signatures as they would need (i) to know the underlying generating \gls{gmm} while only having access to a few samples that change over time; or (ii) a significant computational effort and amount of time to observe the network. Moreover, since signatures are updated every few tens of milliseconds, \veriphy can detect replay attacks by identifying retransmissions of previously used signatures, thus alerting the system regarding ongoing attacks. 

\begin{figure}[!t]
    \vspace{0.2in}
    \centering
    \begin{minipage}{0.28\linewidth}
        \centering
        \includegraphics[width=\linewidth]{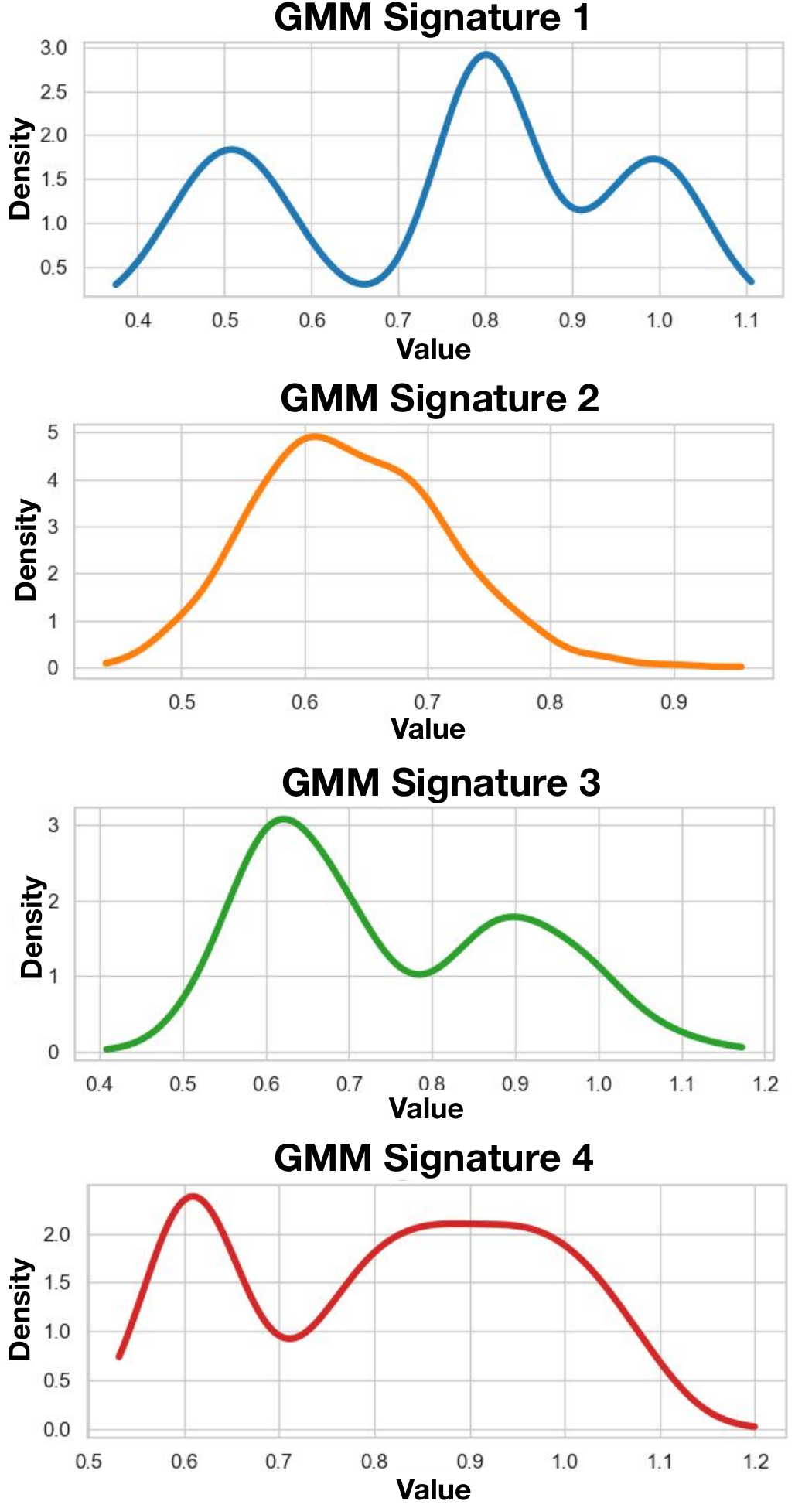}
        \caption{PDFs of four \gls{gmm} signatures.}
        \label{fig:gmm_pdf}
    \end{minipage}%
    \hfill
    \begin{minipage}{0.67\linewidth}
        \centering
        \includegraphics[width=\linewidth]{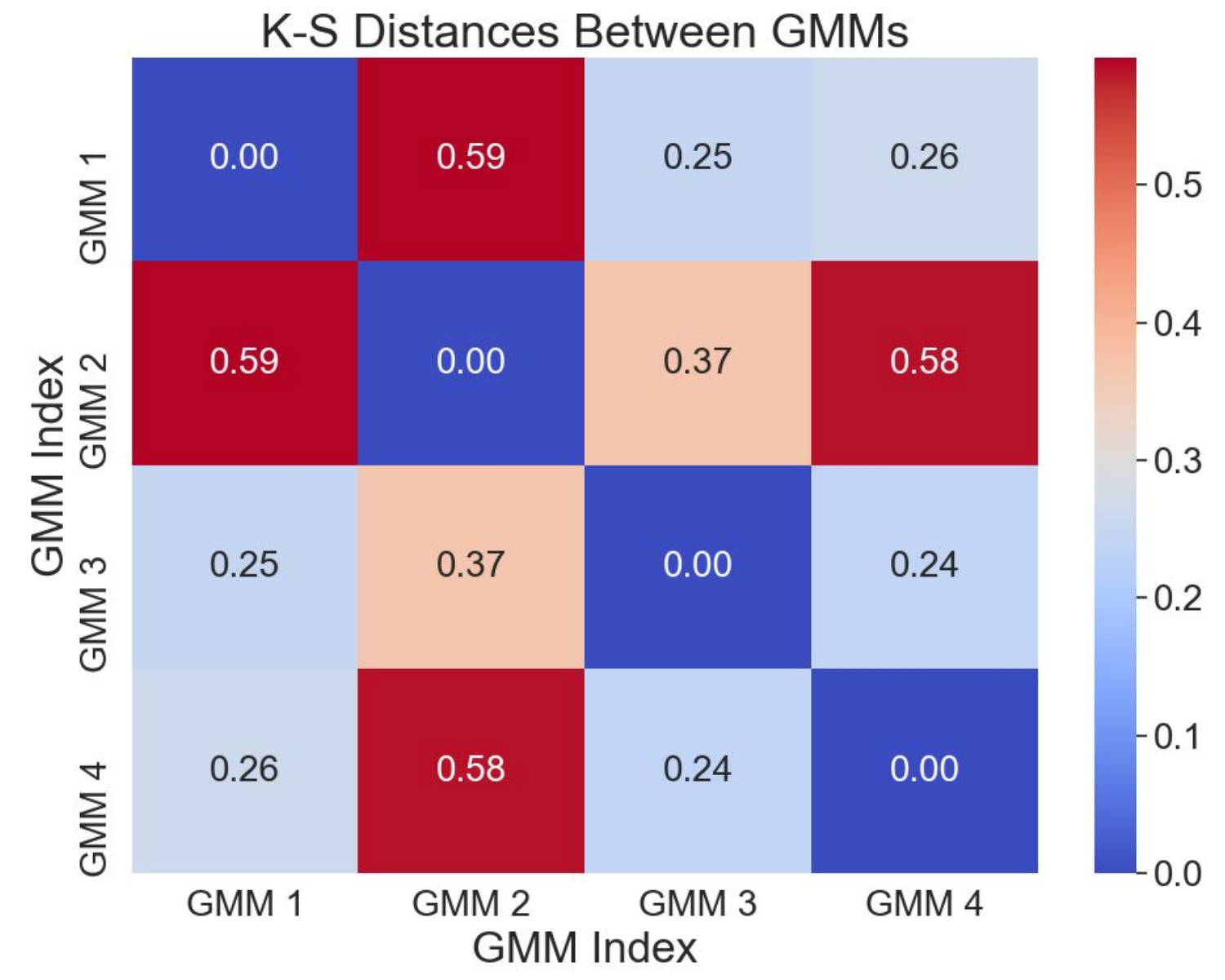}
        \caption{Confusion matrix of pairwise \gls{ks} distances between the four GMMs in Fig.~\ref{fig:gmm_pdf}.}
        \label{fig:ks_dist}
    \end{minipage}
    \vspace{-0.3cm}
\end{figure}

\glsreset{gnb}
\subsection{Signature Detector}
\label{subsec:sig_detect}

The Signature Detector receives I/Q data via an I/Q mirroring technique, which replicates I/Q samples across device modules—such as DU Functions and the Signature Detector—enabling asynchronous processes to run in parallel without service interruption~\cite{cpr_narrow_mirror}. This detector integrates a \gls{dlai}-based \gls{dnn} to identify and verify signatures in the mirrored I/Q samples. As shown in Fig.~\ref{fig:veriphy_main}, I/Q data is simultaneously routed to standard \gls{gnb} functions and the Signature Detector for real-time user authentication.

The Signature Detector flags transmissions based on signature presence. Thus, \veriphy can block authentication requests without valid signatures, preventing unauthorized core network access when stricter security is needed.

\glsreset{gmm}
\glsreset{ks}
\section{Generating and Applying Unique Signatures}
\label{sec:sig_gen}

In our work, we create uniquely identifiable signatures by generating a set of \glspl{gmm} that are significantly different from each other. We achieve this by setting a minimum \gls{ks} distance of $\epsilon$ between each \gls{gmm}. The \gls{ks} distance measures the maximum difference between the \gls{cdf} of these distributions and is an indicator for their similarity. By setting a minimum distance $\epsilon$, we ensure minimal overlap, making each signature distinct. 

\begin{figure}[!b]
    \centering
    \vspace{-0.3cm}
    \includegraphics[width=.9\linewidth]{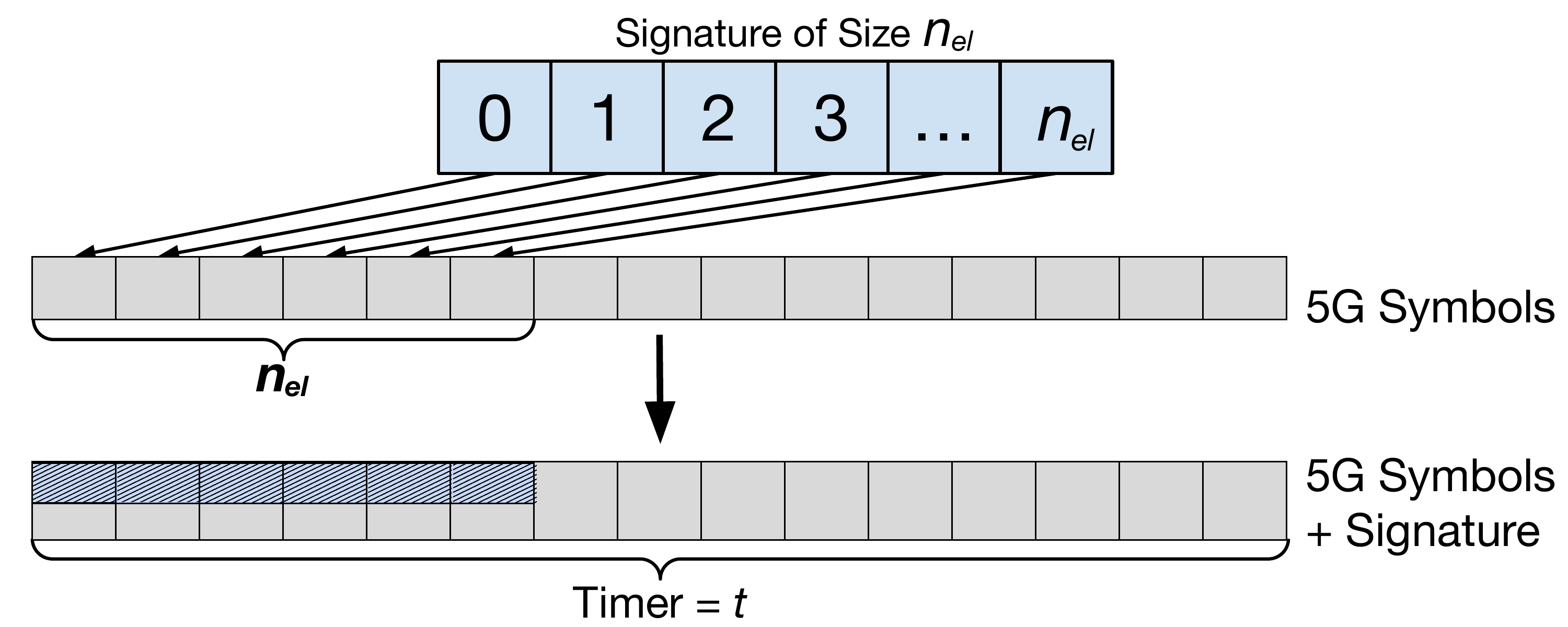}
    \caption{Visualization of how a signature of size $n_{el}$ is sent sequentially over the specified timer $t$.}
    \label{fig:send_sig_seq}
\end{figure}

\subsection{Generating Gaussian Mixture Model Signatures}

To generate uniquely identifiable signatures, \veriphy uses \glspl{gmm} that are sufficiently distinct from one another based on a minimum \gls{ks} distance threshold $\epsilon > 0$. The user specifies the number of signatures $N$ (e.g., number of \glspl{ue}), the perturbation value range $[m_{min}, m_{max}]$, and the maximum number of components $p$ per model. Each \gls{gmm} is generated with randomized peaks and accepted only if its \gls{ks} distance from all previously generated models exceeds $\epsilon$. This ensures that all $N$ signatures are statistically distinguishable, enhancing robustness and minimizing overlap.
Once the \glspl{gmm} are generated, they are exported to the Signature Generators.

For signature generation, we set the \gls{gmm} range to $[0.5, 1]$ to avoid excessive alteration of I/Q samples, leveraging normalization where 1 and 0.5 represent full and half signal levels, respectively. A \gls{ks} distance threshold of $\epsilon > 0.2$ was chosen to ensure sufficient distinctiveness between models while maintaining generation feasibility. It is worth mentioning that higher $\epsilon$ values enforce differentiation across signatures, but might lead to excessive rejections during model creation, making the process longer. However, unless \glspl{gmm} need to be generated in real-time, this problem can be neglected in most scenarios.

Fig.~\ref{fig:gmm_pdf} shows the Power Spectral Densities (PSDs) of four generated \gls{gmm} signatures, each exhibiting a distinct profile. Their pairwise \gls{ks} distances, shown in Fig.~\ref{fig:ks_dist}, range from 0.24 to 0.59, confirming that the signatures are well-separated and sufficiently dissimilar for robust use in \veriphy.

\subsection{Apply Signatures with Wireless Steganography}

To transmit signatures via steganography, a communication channel within the \gls{ue} or \gls{gnb} injects sampled \gls{gmm} distributions on top of user data. Each signature of size $n_{el}$ is sent element by element, followed by a $t$-millisecond delay before the next transmission. Fig.~\ref{fig:send_sig_seq} illustrates this sequence.

To enhance signature uniqueness and concealment, we implement two mechanisms: \textbf{(1)} each transmission uses a newly sampled signature from the \gls{gmm}, making replication possible only for those with access to the model, and \textbf{(2)} randomization and a binary switch determine whether to transmit a signature, preventing pattern inference and reconstruction of the underlying distribution. This ensures each transmission follows a uniquely unpredictable pattern.

\begin{figure}[!t]
    \vspace{0.1cm}
    \centering
    \includegraphics[width=.85\linewidth]{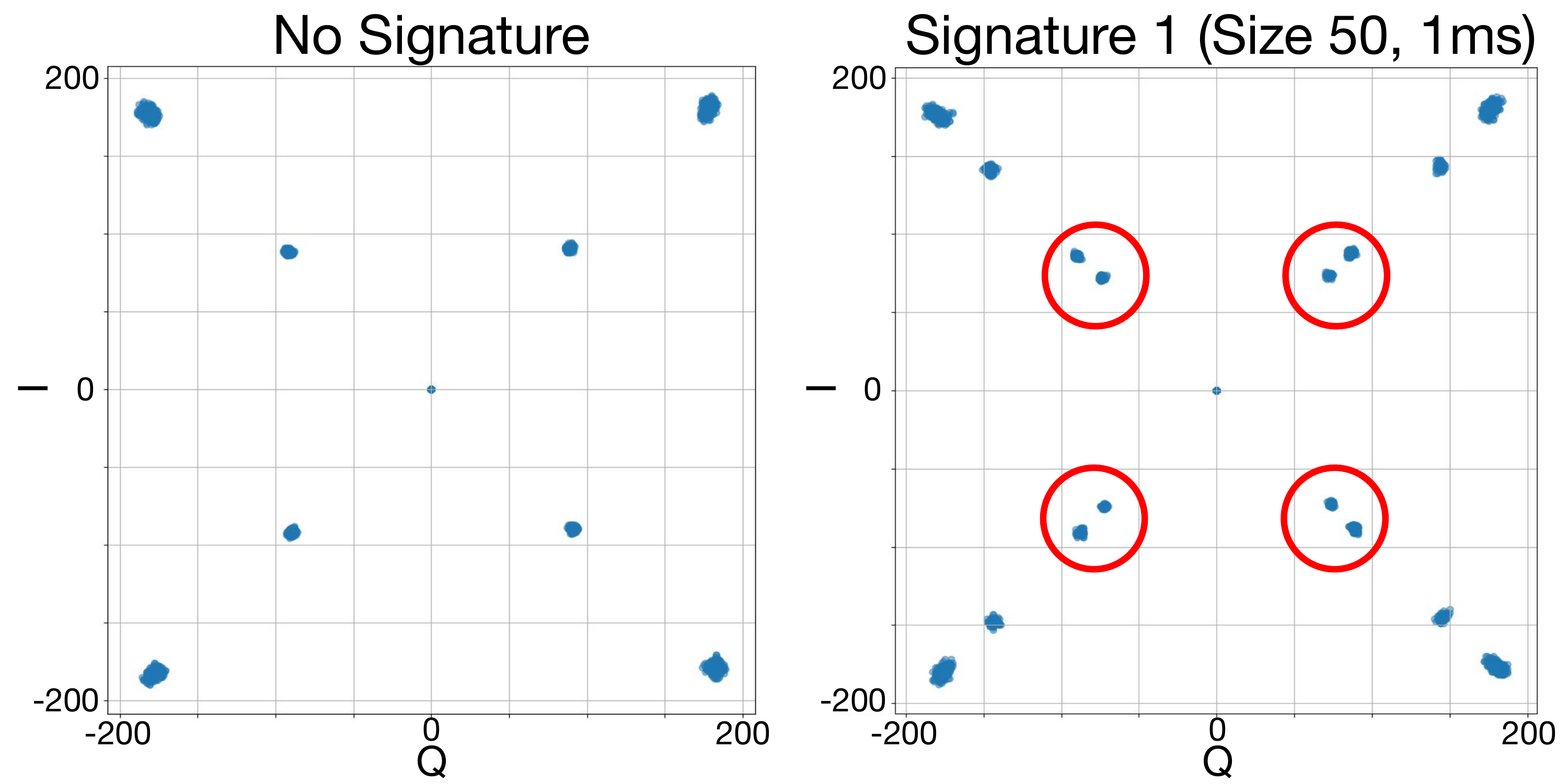}
    \caption{I/Q constellation comparison with and without signature embedding. The left shows the unmodified signal, while the right displays the same signal with an embedded signature (Signature 1, Size 50, Duration 1~ms).} 
    \label{fig:iq_comp}
    \vspace{-0.4cm}
\end{figure}

\begin{figure}[!b]
    \vspace{-0.3cm}
    \centering
    \includegraphics[width=.85\linewidth]{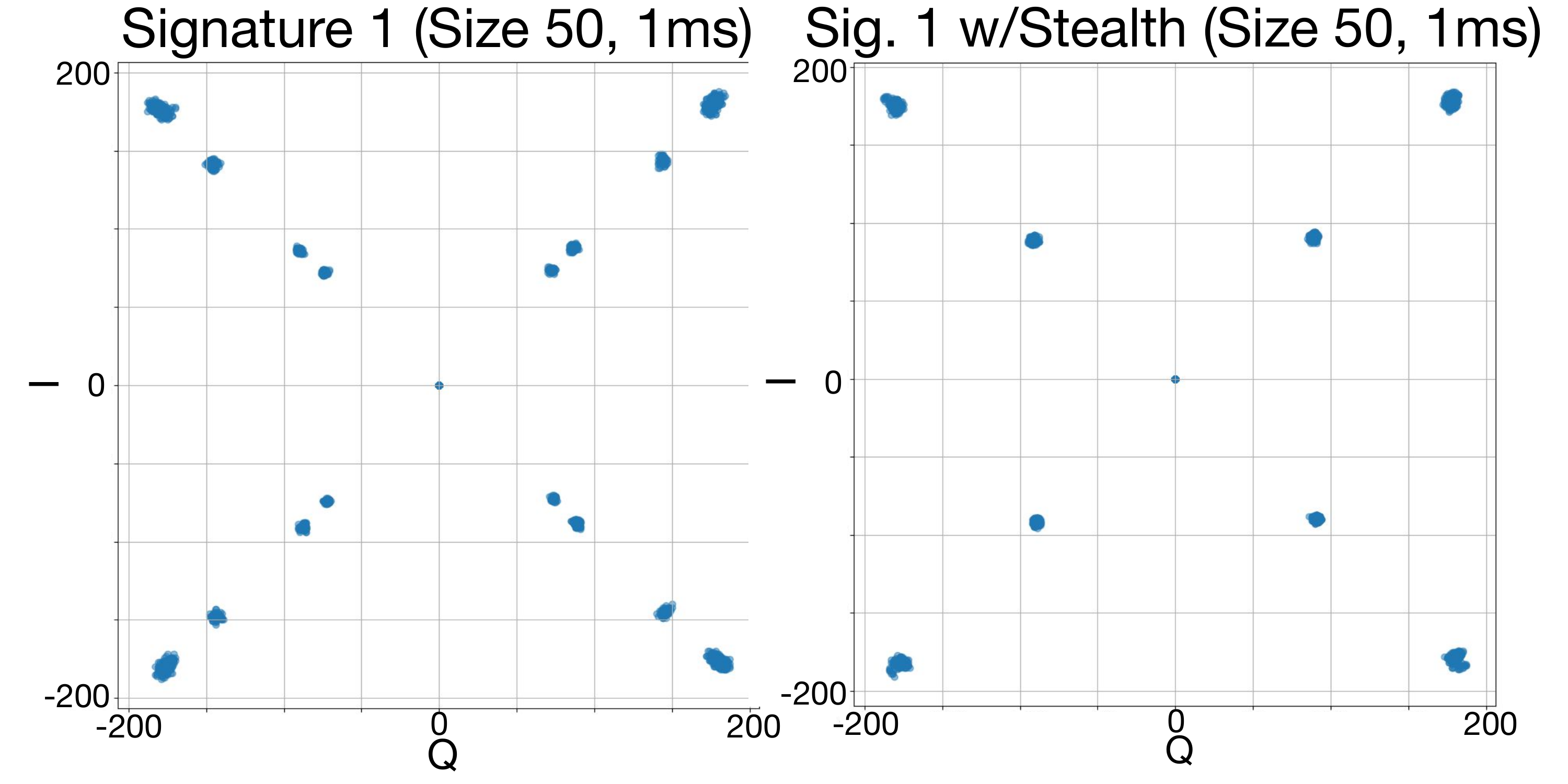}
    \caption{Effect of stealth mode on signal visibility. The left shows a signal with Signature 1 embedded without stealth mode, where the signature is visibly distinguishable. The right shows the same signature with stealth mode applied, significantly obscuring its presence.}
    \label{fig:iq_comp_stealth}
\end{figure}

Transmitting signatures as-is may expose covert activity, as experts could detect anomalies in the I/Q plot. Fig.~\ref{fig:iq_comp} compares 1~ms of normal (left) and signature-embedded (right) transmissions, with red circles highlighting I/Q alterations. This compromises the covertness essential to advanced \gls{pla} systems. To address this, we introduce a stealth mode that preserves signature uniqueness while better concealing their presence.
%
\newline
\textbf{Stealth Mode with Signature Scaling:}
To reduce the detectability of \gls{gmm} signatures, we apply a scaling factor during their generation and embedding, ensuring that modified I/Q samples remain close to the original signal. As shown in Fig.~\ref{fig:iq_comp_stealth}, scaled signatures blend seamlessly into the transmission. During transmission (Fig.~\ref{fig:send_sig_seq}), each I/Q value is scaled to match typical patterns. Without scaling (Fig.~\ref{fig:iq_comp}), signatures are distinct and may reveal user-specific traits. With stealth mode, these differences are minimized, making the signatures nearly indistinguishable from standard traffic, enabling covert yet effective physical-layer authentication.

\begin{figure}[!t]
    \vspace{0.1cm}
    \centering
    \includegraphics[width=.9\linewidth]{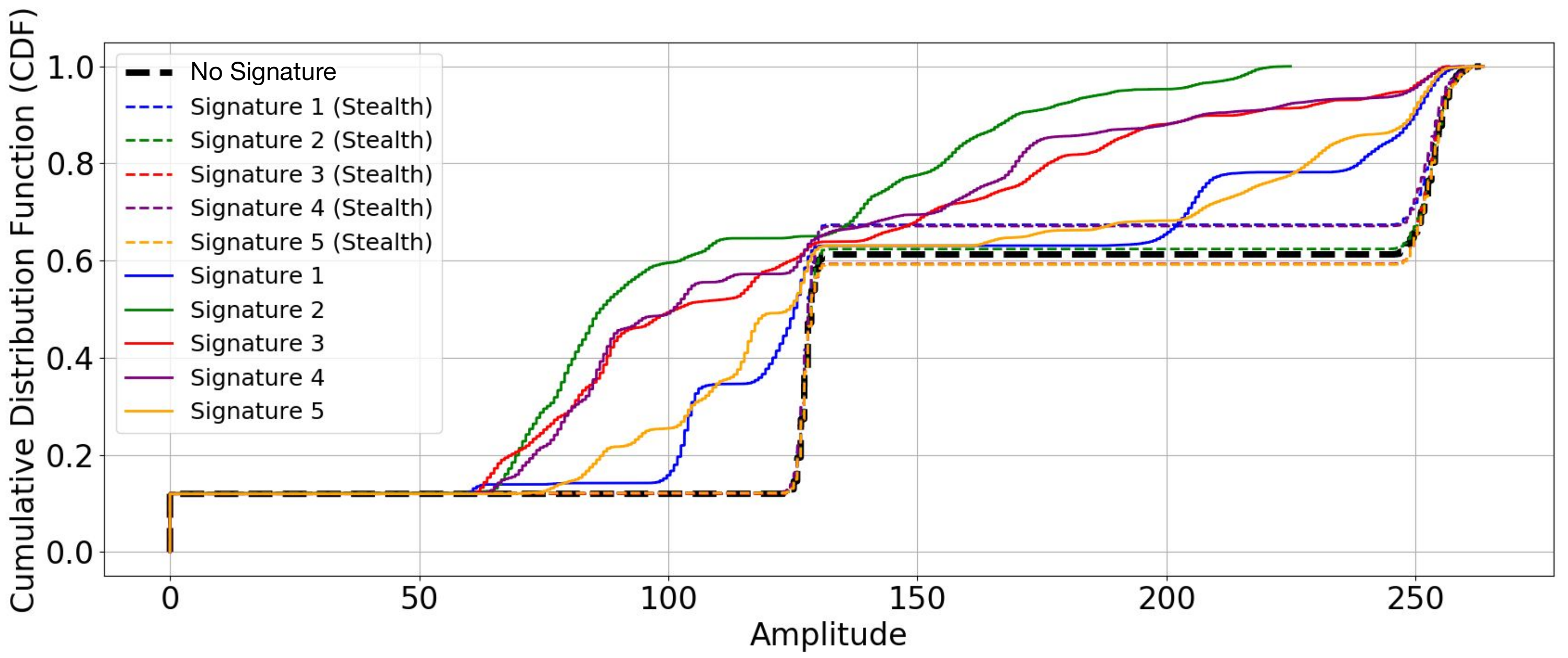}
    \caption{Plot of the Energy CDFs for all signatures, comparing results with and without the use of the signature stealth mode against a transmission without any signal present.}
    \label{fig:cdf}
    \vspace{-0.3cm}
\end{figure}

Another risk to covertness is detection through energy analysis. We compare the Cumulative Distribution Functions (CDFs) of unaltered and scaled \gls{gmm} signatures against standard I/Q samples (no signature) in Fig.~\ref{fig:cdf}. Unaltered signatures show clear deviations from the baseline, revealing energy-level differences. In contrast, scaled signatures closely match the baseline CDF, making them harder to distinguish. This demonstrates that scaling effectively preserves the signal’s energy profile, enhancing stealth and overall security.

In Section~\ref{subsec:stealth}, we will show that we can still guarantee 95\% accuracy even if stealth mode is active.


\section{\textit{VeriPHY} Prototype}
\label{sec:prototype}

We implement our \veriphy prototype on the \gls{oai} RFSim, utilizing its features to simulate realistic 5G network conditions and interactions. To evaluate the capabilities of both the Signature Generator and Detector, we conduct five experiments with varying signature sizes ($n_{el}$) and transmission intervals ($t$), and test two different \glsreset{dnn} models to assess signature detection. Details of our testbed implementation are provided in \ref{subsec:openair}, our \gls{dlai} models are described in \ref{subsec:mt_dl}, and our experimental datasets are discussed in \ref{subsec:datasets}.

\glsreset{oai}
\subsection{OpenAirInterface 5G Platform Implementation}
\label{subsec:openair}

\gls{oai} serves as the implementation environment for demonstrating \veriphy within 5G networks~\cite{nikaein2014openairinterface}. This platform provides an end-to-end, 3GPP-compliant implementation of both the 5G \gls{ran} and core network, making it an ideal choice for prototyping and data collection. By leveraging the \gls{oai} RFSim, which creates a channel with a bandwidth of 40 MHz on \gls{5g} band 78, we can thoroughly test and refine the system’s performance and features in a controlled environment. This flexible setup allows for the integration of various components and testing of different configurations, ensuring that our prototype can adapt to diverse deployment scenarios within 5G networks.

\subsection{Deep Neural Networks for Signature Detection}
\label{subsec:mt_dl}

To implement the Signature Detector, we employ VGG16~\cite{simonyan2014very} and SENet~\cite{hu2018squeeze}. 
Originally designed for image recognition, VGG16 is well-suited for capturing complex 2D I/Q constellation patterns through deep convolutional layers, making it a popular choice for RF signal analysis.
SENet enhances performance by applying channel-wise attention to emphasize key spectral features and is known for low false positive rates in signal classification.
%
We use 2D inputs formed from I/Q groups extracted from \gls{oai}, which stores 60 samples per group (~0.028ms). To analyze 1~ms of data, we require 2160 samples, resulting in an input shape of $(N, 60, 36, 2)$, where $N$ is the number of training examples. This setup aligns with RFSim sampling intervals for accurate 5G signal simulation.

VGG16 is a deep \gls{cnn} with 16 layers and small 3x3 filters, known for strong feature extraction. When adapted for spectrum-based classification with 2D I/Q inputs, VGG16 can recognize complex frequency components and temporal variations, making it well-suited for leveraging spatial and spectral correlations.
Squeeze-and-Excitation Networks (SENet) improvew representational power by modeling interdependencies between channels using a ``squeeze" and ``excitation" mechanism. This aggregates spatial information into channel-wise statistics and dynamically re-weights features, boosting sensitivity to key spectral and temporal patterns. Applied to 2D I/Q inputs, SENet enhances classification accuracy and reduces false-alarms through more precise feature emphasis.

\glsreset{oai}
\subsection{Dataset Generation}
\label{subsec:datasets}

To train our \gls{dlai} models, we generate datasets with enough diversity to ensure accurate training of \glspl{dnn} while avoiding overfitting or underfitting. Using our \gls{oai} implementation, we created five unique datasets, each with different signature sizes and send rates, and five signatures per dataset. The datasets were generated using OAI's RFSimulator to model a 40 MHz bandwidth channel on \gls{5g} band 78. We implemented a system to manage the storage of I/Q data from RFSim, optimizing file size and organization during post-processing. The signatures, generated with \glspl{gmm} to have a \gls{ks} distance of at least 0.2 for uniqueness, were consistent across all datasets, which included the following parameters: \textbf{(1)} $n_{el} = 10$, $t = 1~ms$, \textbf{(2)} $n_{el} = 20$, $t = 1~ms$, \textbf{(3)} $n_{el} = 50$, $t = 1~ms$, \textbf{(4)} $n_{el} = 20$, $t = 20~ms$, and \textbf{(5)} $n_{el} = 50$, $t = 20~ms$. This approach allowed us to test various parameter sets using the same five signatures, ensuring no bias and optimizing the dataset for training.

\section{Experimental Results}
\label{sec:results}

In this section, we present results that illustrate \veriphy effectiveness and accuracy. We begin by profiling the accuracy.  Next, we address latency aspects to demonstrate real-time inference. Finally, we assess the performance of \veriphy when stealth mode is active.


\begin{table}[!b]
\setlength\belowcaptionskip{5pt}
    \centering
    \vspace{-0.3cm}
    \footnotesize
    \setlength{\tabcolsep}{2pt}
    \caption{Accuracy and F-score metrics for the VGG16 and SENet models across different configurations.}
    \label{tab:model_performance}
    \begin{tabularx}{\columnwidth}{
        >{\raggedright\arraybackslash\hsize=0.2\hsize}X 
        >{\raggedright\arraybackslash\hsize=0.4\hsize}X
        >{\centering\arraybackslash\hsize=0.25\hsize}X
        >{\centering\arraybackslash\hsize=0.15\hsize}X}
        \toprule
        Model & Configuration & Accuracy (\%) & F1 Score \\
        \midrule
        VGG16 & Sig. Size: 10, time: 1~ms & 100 & 1.00 \\
              & Sig. Size: 20, time: 1~ms & 99.63  & 1.00 \\
              & Sig. Size: 50, time: 1~ms & 100 & 1.00 \\
              & Sig. Size: 20, time: 20~ms & 57.50 & 0.54 \\
              & Sig. Size: 50, time: 20~ms & 78.05 & 0.78 \\
        \midrule
        SENet & Sig. Size: 10, time: 1~ms & 90.62  & 0.91 \\
              & Sig. Size: 20, time: 1~ms & 89.56  & 0.89 \\
              & Sig. Size: 50, time: 1~ms & 100 & 1.00 \\
              & Sig. Size: 20, time: 20~ms & 100 & 1.00 \\
              & Sig. Size: 50, time: 20~ms & 100 & 1.00 \\
        \bottomrule
    \end{tabularx}
\end{table}

\glsreset{dnn}
\subsection{Model Accuracy}

\begin{figure}[!t]
    \centering
    \begin{subfigure}{0.48\columnwidth}
        \centering
        \includegraphics[width=.88\textwidth]{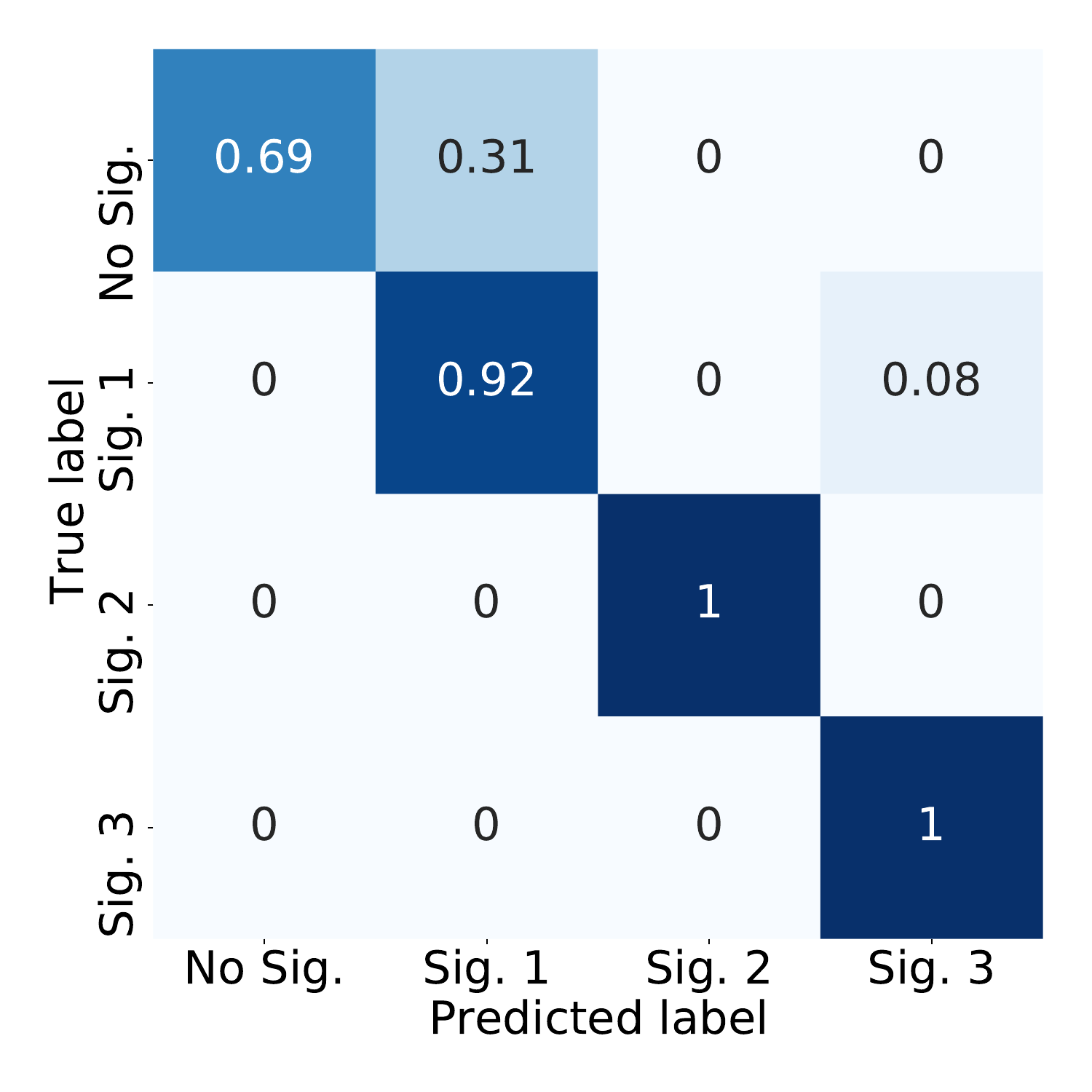}
        \caption{SENet: $n_{el} = 20, t = 1$}
        \label{fig:senet_20_1}
    \end{subfigure}\hfill
    \begin{subfigure}{0.48\columnwidth}
        \centering
        \includegraphics[width=.88\textwidth]{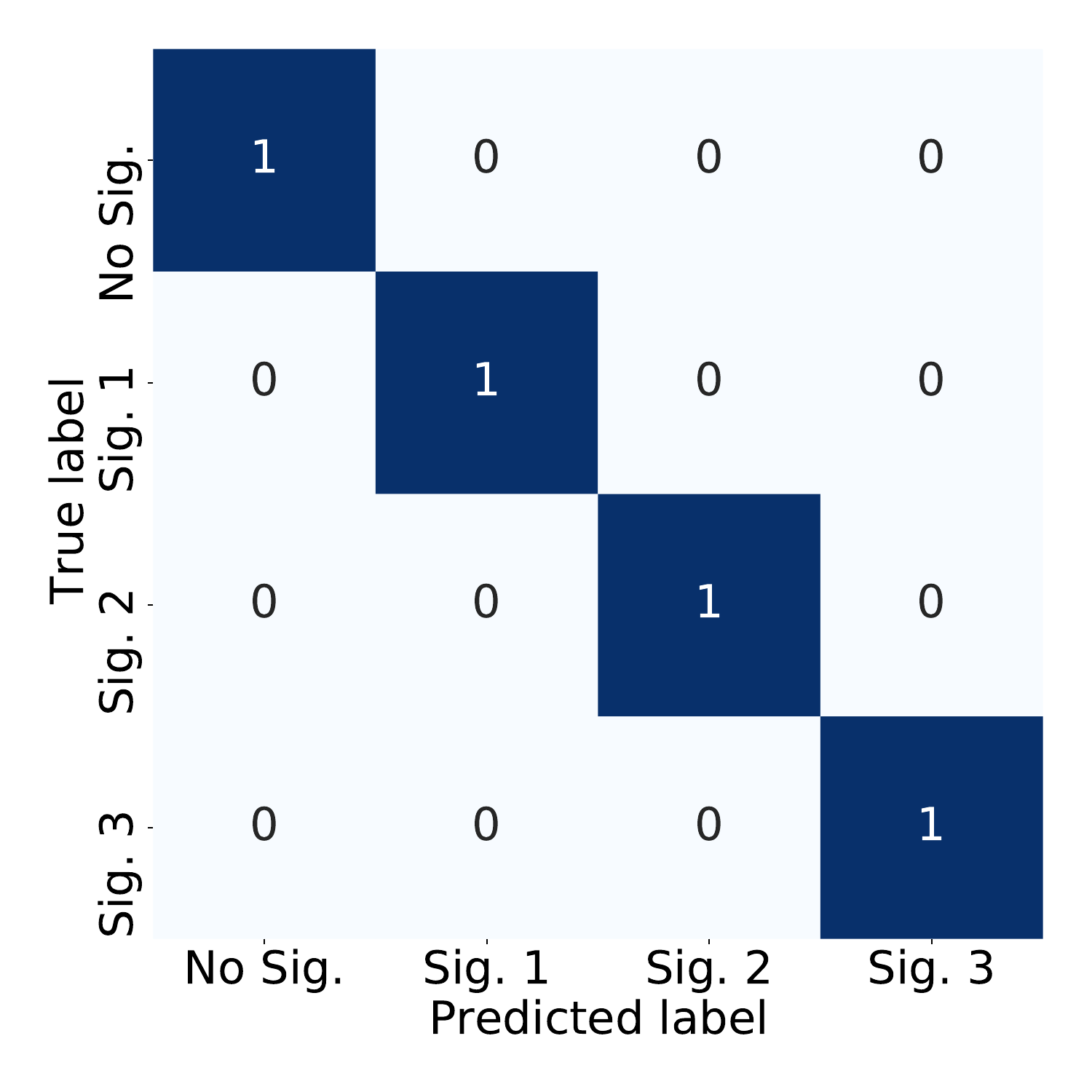}
        \caption{SENet: $n_{el} = 20, t = 20$}
        \label{fig:senet_20_20}
    \end{subfigure}
    
    \vspace{1em}
    
    \begin{subfigure}{0.48\columnwidth}
        \centering
        \includegraphics[width=.88\textwidth]{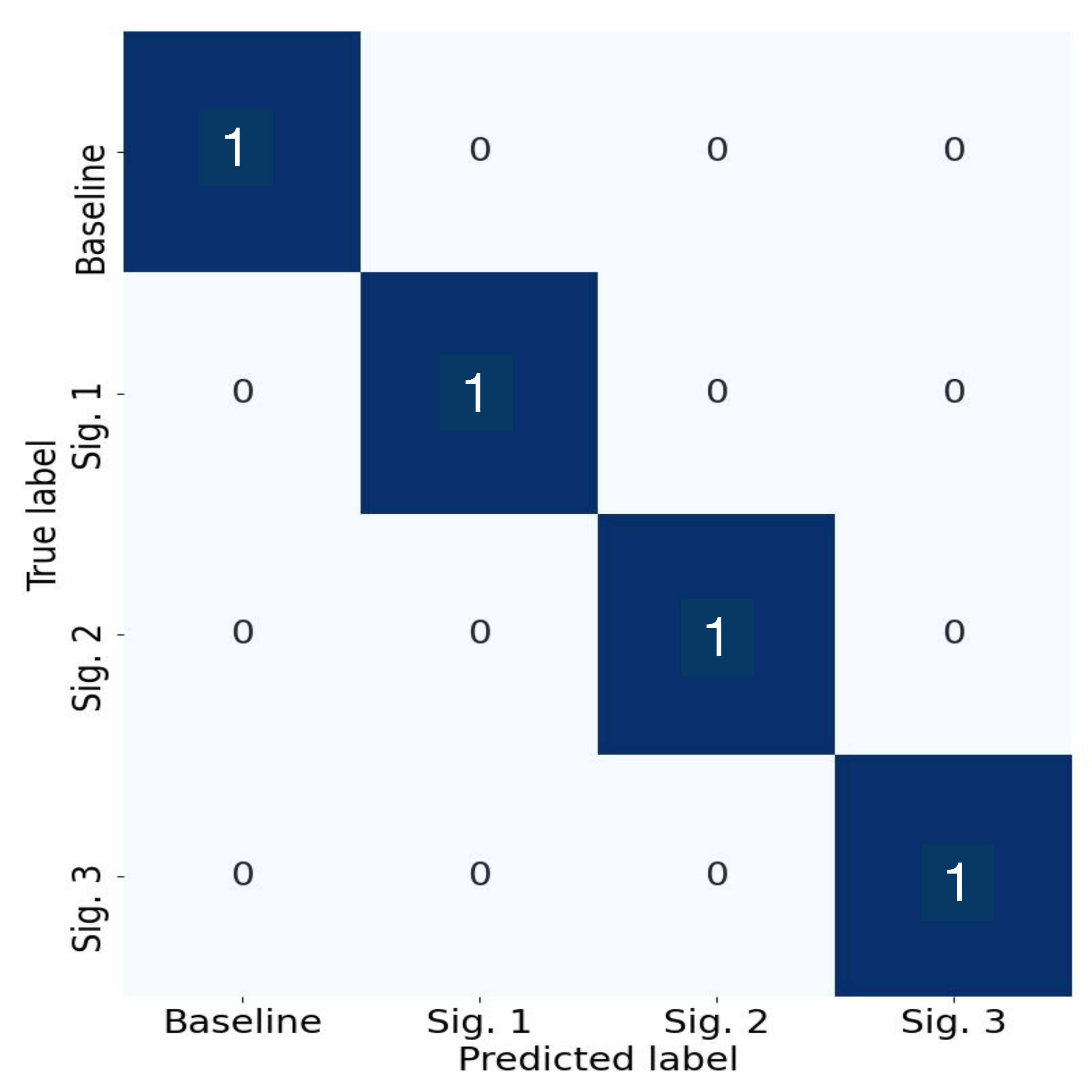}
        \caption{VGG16: $n_{el} = 50, t = 1$}
        \label{fig:vgg_50_1}
    \end{subfigure}\hfill
    \begin{subfigure}{0.48\columnwidth}
        \centering
        \includegraphics[width=.88\textwidth]{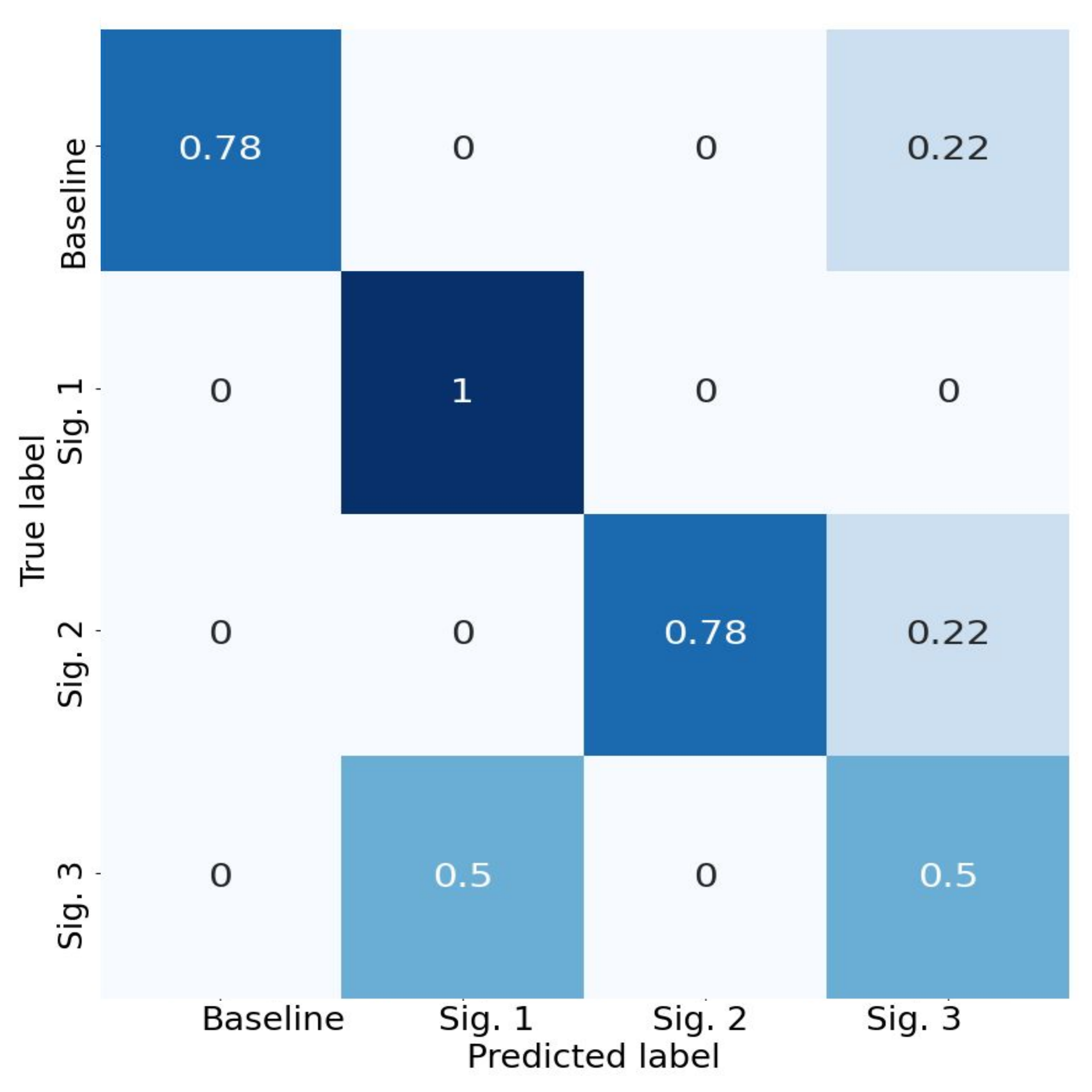}
        \caption{VGG16: $n_{el} = 50, t = 20$}
        \label{fig:vgg_50_50}
    \end{subfigure}
    
    \caption{Confusion Matrices for four of the trained models across different experimental setups. \textit{(All times, $t$, reported in milliseconds)}}
    \label{fig:cm_group}
    \vspace{-0.3cm}
\end{figure}

For both VGG16 and SENet, five models were trained across different signature deployments. Three tests used signatures sent every 1ms with $n_{el}$ sizes of 10, 20, and 50, while two tests used 20ms intervals with sizes 20 and 50. These configurations are summarized in Table~\ref{tab:model_performance}.
The VGG16 model performs well with 100\% accuracy and F1 scores of 1.00 for 1ms signatures with sizes 10 and 50 (Fig.~\ref{fig:vgg_50_1}). However, performance drops when signatures are sent every 20ms, with accuracy falling to 57.50\% and 78.05\% for sizes 20 and 50, respectively (Fig.~\ref{fig:vgg_50_50}), indicating that the model struggles when the signature becomes a smaller portion of the longer I/Q data.

The SENet model shows strong performance with some variability depending on the configuration. For 1ms signatures, sizes 10 and 20 achieve accuracies of 90.62\% and 89.56\% (Fig.~\ref{fig:senet_20_1}), with F1 scores of 0.91 and 0.89. However, it excels in more complex scenarios, achieving 100\% accuracy and an F1 score of 1.00 for signature size 50 at 1ms, and for 20ms signatures of sizes 20 and 50 (Fig.~\ref{fig:senet_20_20}).

VGG16 performs well with fast signatures (1ms) but has lower performance at 20ms. SENet instead delivers consistent performance across all configurations, making it a more adaptable choice for \veriphy.


\vspace{-0.2cm}
\subsection{Model Latency}


The latency values for the VGG16 and SENet models across varying signal sizes and send intervals demonstrate their computational efficiency. For VGG16, latency remains consistent between 7.36 and 7.41~ms when the send interval is 1~ms, with a slight increase to 7.79~ms at a 20~ms interval. SENet shows similar stability, maintaining latency between 5.29 and 5.36~ms at 1~ms, and rising modestly to 5.32–5.44~ms at 20~ms. Both models handle changes in signal size well, though longer send intervals lead to minor latency increases.

The SENet model consistently shows lower latency than VGG16 across all conditions, suggesting better efficiency for quick-response tasks. Both models exhibit minimal sensitivity to signal size changes but show a slight latency increase with longer send times, highlighting the importance of optimization for real-time applications.

\begin{table}[!b]
\setlength\belowcaptionskip{5pt}
    \centering
    \vspace{-0.3cm}
    \footnotesize
    \setlength{\tabcolsep}{2pt}
    \caption{Inference times for models with different time configurations.}
    \label{tab:inference_times}
    \begin{tabularx}{\columnwidth}{
        >{\raggedright\arraybackslash\hsize=0.42\hsize}X 
        >{\centering\arraybackslash\hsize=0.29\hsize}X
        >{\centering\arraybackslash\hsize=0.29\hsize}X}
        \toprule
        Metric & 1~ms Configuration & 20~ms Configuration \\
        \midrule
        I/Q Capture Time & 1.00~ms & 20.00~ms \\
        Average Processing Time & 0.037~ms & 0.038~ms \\
        Average CNN Input Time & 0.142~ms & 0.147~ms \\
        \midrule
        VGG16 Model Latency  & 7.377~ms & 7.712~ms \\
        \textbf{VGG16 Total Inference Time} & \textbf{8.562~ms} & \textbf{27.898~ms} \\
        \midrule
        SENet Model Latency & 5.316~ms & 5.378~ms \\
        \textbf{SENet Total Inference Time} & \textbf{6.501~ms} & \textbf{25.564~ms} \\
        \bottomrule
    \end{tabularx}
    \vspace{0.3cm}
\end{table}


\subsection{\veriphy Inference Time}

Table~\ref{tab:inference_times} shows the inference times for VGG16 and SENet models across all configurations, highlighting the acquisition of signals (I/Q Capture Time), processing, and CNN input times.

In the 1~ms configuration, VGG16 has a total inference time of 8.562~ms, with 1.00~ms for I/Q capture, 0.037~ms for processing, 0.142~ms for CNN input generation, and 7.377~ms of model latency. In contrast, SENet's total inference time is 6.501~ms, with a model latency of 5.316~ms.
For the 20ms configuration, VGG16's total inference time increases to 27.898~ms due to the 20.00~ms I/Q capture time, while processing and CNN input generation remain stable. The model latency slightly rises to 7.712~ms. Similarly, SENet's total inference time increases to 25.564~ms, with model latency at 5.378~ms.



Analyzing the total inference time across signature intervals reveals an important efficiency trade-off. While the 20~ms interval introduces some delay compared to the 1~ms interval, it allows decisions to be made before the I/Q buffer is fully utilized, reducing latency from buffer processing. Therefore, while the 1~ms model has faster inference times, the 20~ms deployment may offer better overall efficiency by optimizing data throughput and minimizing buffer delays.

\subsection{Normal Mode vs Stealth Mode}
\label{subsec:stealth}

To evaluate the impact of stealth mode on \veriphy signature detection, we compared detection accuracy using SENet. For a stealth model with $n_{el} = 50$ and $t = 1~ms$, accuracy drops slightly by 3\% to 97\%, indicating stealth mode does not significantly disrupt detection.



\begin{figure}[!t]
    \vspace{0.1cm}
    \centering
    \begin{subfigure}{0.48\columnwidth}
        \centering
        \includegraphics[width=.88\textwidth]{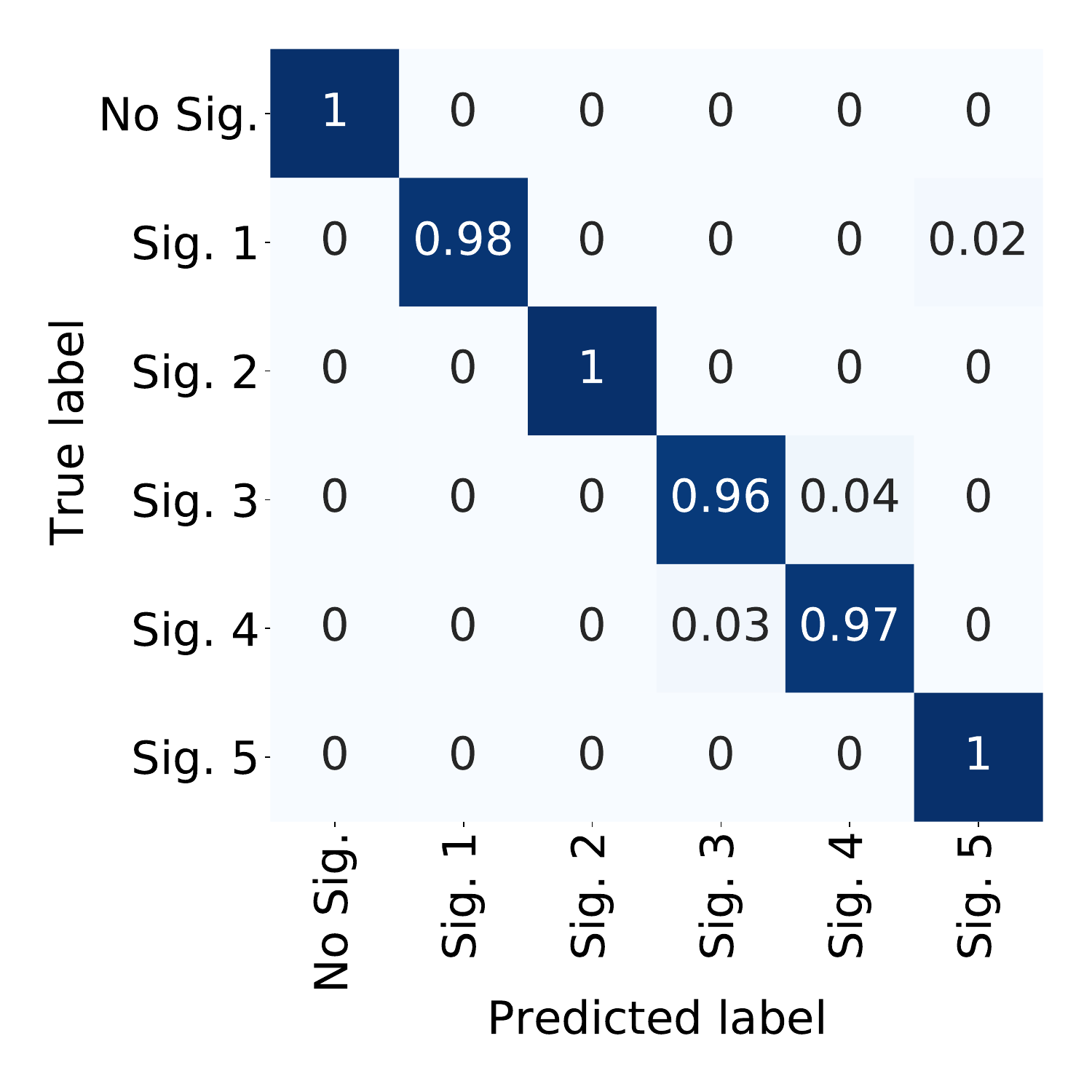}
        \caption{SENet, 5 Signatures (Normal)}
        \label{fig:sub_senet_6}
    \end{subfigure}\hfill
    \begin{subfigure}{0.48\columnwidth}
        \centering
        \includegraphics[width=.88\textwidth]{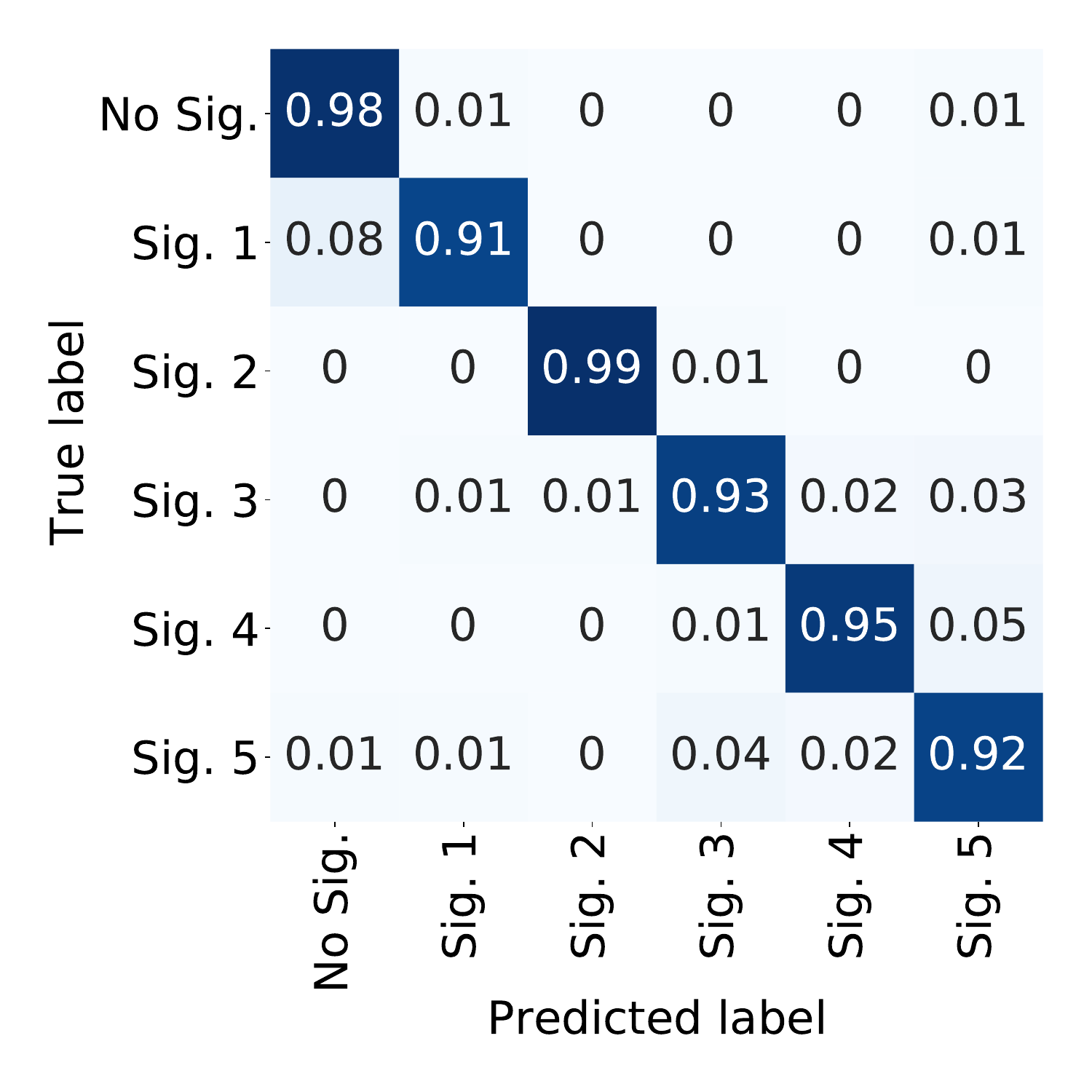}
        \caption{SENet, 5 Signatures (Stealth)}
        \label{fig:sub_stealth_6}
    \end{subfigure}
    \caption{Confusion matrices: normal vs. stealth \veriphy ($n_{el}=50$, $t=50$~ms).}
    \label{fig:stealth_comp_cm}
    \vspace{-0.3cm}
\end{figure}

To further evaluate our stealth mode, we introduced two additional signatures and trained both a standard and a stealth-enhanced model using $n_{el} = 50$ and $t = 1~ms$. As shown in Fig.~\ref{fig:stealth_comp_cm}, there is a slight drop in accuracy from 98.5\% in the normal model to 94.6\% in the stealth model. Specifically, the accuracy for `Sig. 1' drops from 0.98 to 0.91 (a 7\% decrease), and for `Sig. 4' from 0.97 to 0.95 (a 2\% decrease). Despite this, the stealth model maintains strong classification performance. These results indicate that while stealth characteristics introduce minor degradation, the model remains highly effective and resilient, successfully generalizing to the altered signature patterns with minimal impact on accuracy.



\glsreset{pla}
\section{Conclusion}
\label{sec:conclusion}

In this paper, we proposed \veriphy, a novel \gls{dlai}-based \gls{pla} solution for 5G networks that embeds device signatures into I/Q transmissions using steganography. \glspl{gmm} generate pseudo-random, uniquely identifiable signatures with distinctiveness ensured by \gls{ks} distance, complicating replication. Trained DNNs achieved 93–100\% detection accuracy with latencies as low as 6.5~ms. We also introduced a \textit{stealth mode} that conceals signatures 
while preserving over 93\% detection accuracy.



\balance
\footnotesize
\bibliographystyle{IEEEtran}
\bibliography{bib}

\end{document}